# Colloidal Quantum Dot Molecules Manifesting Quantum Coupling at Room Temperature

Jiabin Cui[1,2,†], Yossef E. Panfil[1,2,†], Somnath Koley[1,2,†], Doaa Shamalia[1,2], Nir Waiskopf[1,2], Sergei Remennik[2], Inna Popov[2], Meirav Oded[1,2] & Uri Banin[1,2]*

1 Institute of Chemistry, The Hebrew University of Jerusalem, Jerusalem 91904, Israel.
2 The Center for Nanoscience and Nanotechnology, The Hebrew University of Jerusalem, Jerusalem 91904, Israel.
† These authors contributed equally to this work.
* Corresponding author. Email: uri.banin@mail.huji.ac.il (U.B.)

## Abstract

Coupling of atoms is the basis of chemistry, yielding the beauty and richness of molecules. We utilize semiconductor nanocrystals as artificial atoms to form nanocrystal molecules that are structurally and electronically coupled. CdSe/CdS core/shell nanocrystals are linked to form dimers which are then fused via constrained oriented attachment. The possible nanocrystal facets in which such fusion takes place are analyzed with atomic resolution revealing the distribution of possible crystal fusion scenarios. Coherent coupling and wavefunction hybridization are manifested by a red shift of the band gap, in agreement with quantum mechanical simulations. Single nanoparticle spectroscopy unravels the attributes of coupled nanocrystal dimers related to the unique combination of quantum mechanical tunneling and energy transfer mechanisms. This sets the stage for *nanocrystals chemistry* to yield a diverse selection of coupled nanocrystal molecules constructed from controlled core/shell nanocrystal building blocks. These are of direct relevance for numerous applications in displays, sensing, biological tagging and emerging quantum technologies.

# Introduction

Colloidal semiconductor Quantum Dots (CQDs) that contain hundreds to thousands of atoms have reached an exquisite level of control, side by side with gaining fundamental understanding of their size, composition and surface controlled properties leading to their implementation in technological applications [1]. The strongly quantum confined energetic levels of CQDs possess atomic like character, for example - *s* and *p* states, related to their spherical symmetry. This, alongside with the ability to manipulate CQDs into more elaborate structures, naturally led to their consideration as "artificial atoms". Inspired by molecular chemistry, in which functionality of molecules depends on how atoms couple, we apply analogous concepts to enrich CQDs based materials. If one considers CQDs as artificial atom building blocks [2,3], how plentiful would be the selection of composition, properties and functionalities of the corresponding artificial molecules? Herein we introduce the utilization of CQDs as basic elements in "nanocrystal chemistry" for construction of coupled colloidal nanocrystals molecules focusing on homodimer quantum dots (QDs), in analogy to homonuclear diatomic molecules.

Coupled quantum dots were prepared by means of molecular beam epitaxy (MBE) [4–6]. However, MBE-grown double quantum dot structures exhibit some limitations. First, the size of MBE-grown QDs is larger than the colloidal ones, and the typically large distance between the QDs limits wave-function tunneling that yields coupling phenomenon. Correspondingly, such structures exhibit wave-function tunneling that typically yields coupling energies of a few meV confining their utility to low temperature operation in specialized cryogenic applications [7,8]. Furthermore, MBE grown structures are inherently buried within a host semiconductor[9]. In contrast, colloidal quantum dots are free in solution and accessible for wet-chemical manipulations through their surface functionalization. Using such knobs, CQD molecules were constructed by connection with DNA strands providing geometrical control [10], yet in such structures the linker DNA molecules form a barrier that minimizes quantum mechanical coupling. Addressing this limitation, core/multishell structures with concentric regions were first examples of coupling within CQDs architectures, where the wave-functions of two well regions within such NCs may interact leading to CQDs showing dual emission peaks [11]. Other examples constitute synthesis of dot-in-rod structures and growing an additional quantum dot region on the rod apex thus yielding a coupled system [12], and dumbbell architectures [13]. However, these progresses were either restricted by specific morphologies [14], specific materials and relatively large coupling barrier distance and height [15–17]. Therefore, there is a lack of a general approach for producing coupled CQD molecules in which there is flexibility to tailor the potential energy landscape and to tune the coupling strength.

To this end we introduce a facile and powerful strategy for coupled CQD molecules with precise control over the composition and size of the barrier in between them to allow for tuning their electronic coupling characteristics and optical properties. This entails the use of core/shell CQDs as artificial atom building blocks. In terms of the band gap engineering, in first instance, tuning the core size is used to manipulate the wave-functions and energies of the electron and hole. On top of this, further control is afforded by the synthesis of shells on these cores.

While the chemical bond is the basis for combining atoms in molecules, connecting CQDs has to occur through adjoining of their crystal faces to form a continuous crystal. Thus, fusing two core/shell CQDs yields a homodimer with a tailored barrier dictated by the shell composition, thickness and fusion reaction conditions. With such control, using high resolution aberration corrected scanning transmission electron microscopy, we observed and analyzed the orientation relationships including homo-plane-attachment and hetero-plane-attachment in the fusion process. Moreover, the manifestations of quantum coupling were revealed by the broadening and red shift of the band gap transition observed in absorption and photoluminescence, in agreement with the quantum-mechanical calculations for the system. The scale of the hybridization energies and corresponding shifts are significantly lower as compared to diatomic molecules. This is expected considering the much larger dimensions of the CQD building blocks compared to atoms and the different potential energy landscape within the CQD molecule. The coupling also leads to broadening of the excited state transitions of the CQD dimers and the absorption spectrum for the high energy bands is modified as well. The emerging attributes of coupling are also revealed by single nanoparticle spectroscopy studies yielding modified electron-hole recombination rates and single photon statistics in CQD dimers in comparison to monomers.

The approach introduced herein, serves as a basis for a wide selection of CQD molecules utilizing the rich collection of the artificial atom core/shell CQD building blocks. Such CQD molecules bear significant promise for their utilization in numerous applications, including in light-emitting devices, displays, photovoltaics and sensors. For example, the controlled formation of heterodimers consisting of CQD monomers with varying core sizes is of direct relevance for dual color emission. Similarly, forming a heterodimer with a staggered (type-II) band alignment between the two CQD building blocks is envisioned for electric field sensing. Additionally, there is high potential for CQD molecules to be used in emergent quantum technologies such as quantum computation[8]. MBE grown QD molecules were already demonstrated as quantum gates utilizing the interaction between the quantum dots through tunneling [5]. CQD molecules offer enhanced coupling efficiency by their smaller sizes and by the small distances between the QDs resulting in quantum mechanical coupling an order of magnitude larger than prior MBE grown QD systems that is well resolved even at room temperature. This advance significantly widens the scope of quantum technologies applications of coupled quantum dot systems.

## Results

**Formation of coupled CQDs molecules**

Exemplary coupled homodimer molecules were generated from CdSe/CdS core/shell [18,19] CQDs via a procedure utilizing silica nanoparticles as a template for forming molecularly linked dimers [20], which are then fused via a high temperature reaction (Fig. 1a, full scheme in Supplementary Fig.1). Three different CdSe/CdS core/shell CQDs were studied (1.9/4.0 nm, 1.4/2.1 nm, and 1.2/2.1 nm core-radius/shell-thickness, see SI materials and methods for synthesis details and Supplementary Fig.2 for transmission electron microscopy (TEM) images and optical spectra). The TEM and high angle annular dark field (HAADF) scanning transmission electron microscopy (STEM) characterization manifests the wurtzite structure of the monomer CdSe/CdS QDs (Supplementary Fig.3). These CdSe/CdS CQDs were bonded via thiol linking to the surface of a $SiO_2$ nanoparticle template substrate (Supplementary Fig.4-5). A second $SiO_2$ layer was grown for masking the remaining $SiO_2$ surface and to immobilize the bonded CQDs (Supplementary Fig.6), followed by treatment with a tetra-thiol linker (Supplementary Fig.7). Adding a second CQD leads to formation of a molecularly linked dimer structure (Fig. 1b, Supplementary Fig.8). Next, the $SiO_2$ template nanoparticles were selectively etched by HF treatment. Size-selective precipitation was used to separate out the monomers and obtain a highly dimer enriched sample (Fig. 1c).

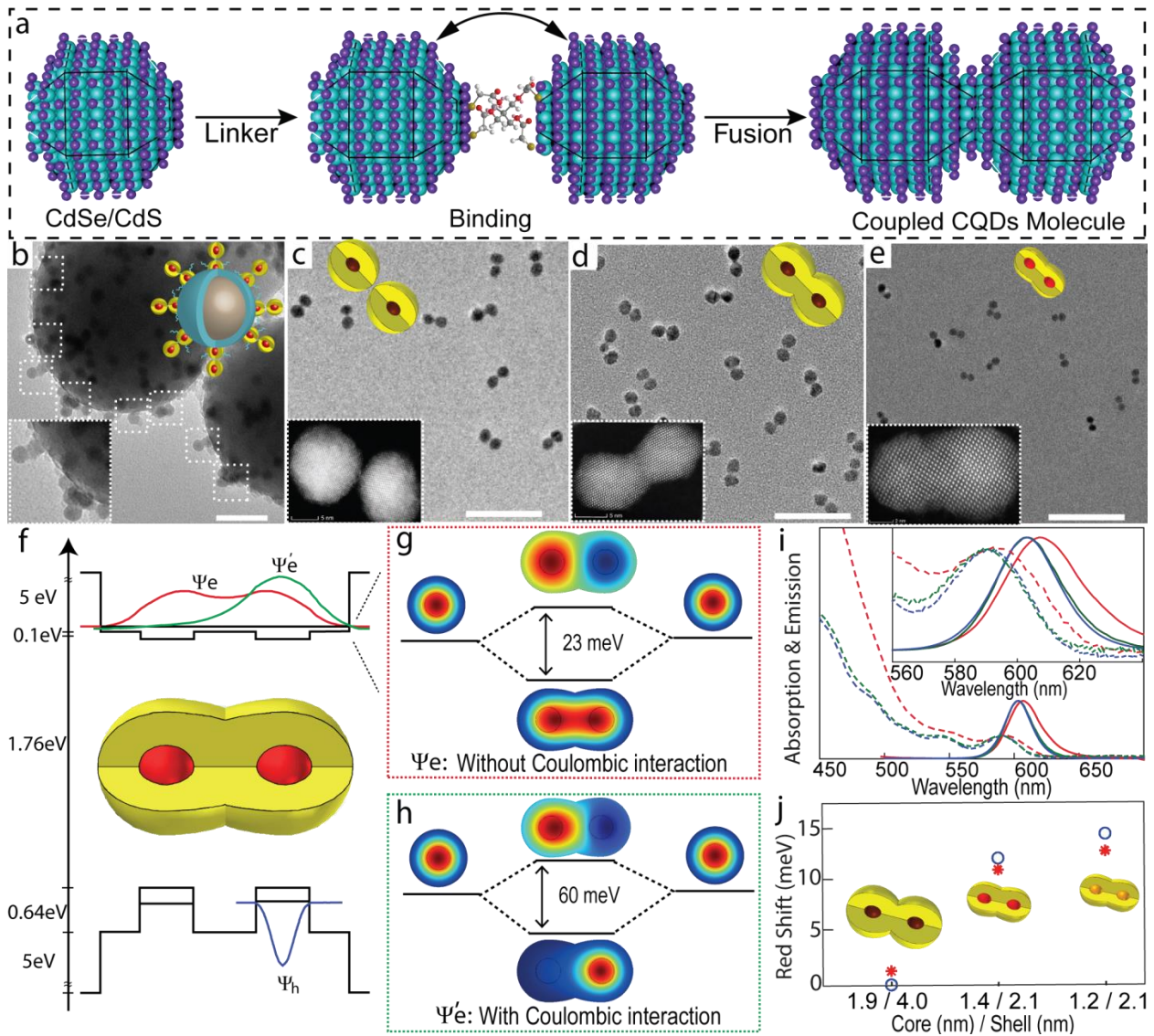

**Fig. 1. Coupled CQDs molecules.** (a) Scheme for fabrication of coupled CdSe/CdS CQD molecule. (b) The dimer@SiO$_2$ CQD structure. The dimer 1.9/4.0 nm CQD molecules (c) before, and (d) after the fusion procedure. (e) The 1.4/2.1 nm fused CdSe/CdS CQD molecules. Schematic structures are illustrated. Scale bars (b-e) are 50 nm and insets show higher magnification images. (f) The potential energy landscape and a cross-section of the calculated first electron wave-function without Coulombic interaction $\Psi_e$ (red), with Coulombic interaction $\Psi'_e$ (green) and of the hole wave-functions $\Psi_h$ (blue) of the coupled CQD molecules. (g) Calculated bonding and antibonding 2-dimensional electron wave-functions without (cross-section of the bonding state is the red curve in f), and (h) with Coulombic interaction (cross-section of the bonding state is the green curve in f). (i) Absorption (dashed lines) and fluorescence spectra (solid lines) of monomers (blue), unfused (green), and fused 1.4/2.1 nm CdSe/CdS CQD molecules (red). (j)
5

Calculated (red asterisk) and experimental (blue circles) band gap red shift of monomer-to-respective-homodimer structures for CQD molecules with different core/shell dimensions.

The dimerization procedure yields a dimer structure with an organic insulating barrier. Hence, to achieve a coupled system, a last step of fusion is required. The fusion procedure was performed while adding Cd-oleate and heating to 180 °C for 20 h. Fig. 1d-e presents the fused dimer structure after a size selection procedure (Supplementary Fig.9-10). At this non-trivial important stage, the reaction parameters, including temperature, time and ligands type and concentration, have a significant influence on the coupled dimers formation. If the temperature was too high (above 240 °C), collapse of the dimer structures through linker bond cleavage may occur, as well as CQD ripening distorting the core/shell architectures. On the other hand, if the temperature was too low, the fusion rate would be too slow and inefficient. The dimer structure formation is also very sensitive to excess of ligands in the solution, which inhibits the fusion and leads to a decrease in the dimer yield. Therefore, careful tuning and choice of these reaction parameters is crucial for achieving high dimer yields and lower yields of dimer collapse and ripening, while achieving a continuous linking region of the shell materials forming the barrier between the two cores in the fused dimers. Further considerations of this important fusion step and the resultant interfacial structures are discussed later.

**Optical signatures for coupling and wavefunction hybridization**

After the fusion step, the resultant CQD dimer leaves an interesting optical signature of a red-shift in the absorption and photoluminescence spectra along with broadening of the band gap and excited state spectral features (Fig. 1 and Supplementary Fig. 11). Generally, there are several factors which can lead to a red-shift: the formation of alloying shell [21], alteration of the dielectric environment [22] (surface ligands) or interfacial strain [23]. To address these different possibilities, we also studied the spectral properties of the monomers, which underwent the fusion reaction under similar conditions (Supplementary Fig. 9-10), and found them to be identical to the original monomer particles (Supplementary Fig. 12). Hence the possibility of observing a red shift in the band gap transition due to formation of an alloy shell or altered dielectric environment can be ruled out. Furthermore, strain effects and change in the dielectric properties during the fusion procedure can be considered negligible as we did not grow an additional shell, but rather the fused shell material is the same (CdS), and the surface ligands are also the same for the CQDs monomers and dimers. Moreover, no shift was observed after the fusion of the large 1.9/4.0nm CQDs (Fig.1), where dielectric and strain effects, if significant, would be expected to contribute as well. In fact, the red shift in the band gap transitions was found to depend systematically on the alteration of the core size and shell thickness of the monomer counterparts, increasing for small core and shell dimensions (Fig 1j). This is consistent with the difference in the delocalization of the wavefunctions in the various CQDs that lead to different degree of coupling of the corresponding wavefunctions in the CQD molecules.

To this end, we have employed quantum mechanical calculations to visualize the wavefunction hybridization and to calculate the expected red-shift in the different CQD dimers. The changed potential energy landscape upon fusion leads to hybridization of the monomer QD wave-functions in the dimers (Fig. 1f), in analogy to homonuclear diatomic molecules. We utilized finite element software (COMSOL) to calculate the energy levels and wave-functions of the fused CdSe/CdS dimer and monomers within an effective-mass based approximation (Supplementary Discussion and Supplementary Table 1 for details). The conduction band in this system is demonstrating the fundamental textbook example of hybridization. According to this model, when the distance between two atoms is decreased, their wave-functions will hybridize to form a symmetric bonding state and anti-symmetric anti-bonding state with energy difference of twice the hopping energy. The bonding and anti-bonding electron wave-functions, which, respectively, are in-phase and anti-phase superpositions of the monomer wave-functions, are presented in Fig.1g for the case of 1.4 nm core radius and a potential energy barrier between the dots of 4.2 nm (0.1eV band-offset), corresponding to the CQD molecules formed from 1.4/2.1 nm core/shell CQDs. Because of the quasi-type II nature of the CdSe/CdS interface, the monomers electron wave-functions are easily hybridized and leading to 23 meV energy spacing between the bonding and anti-bonding electron states. For the hole however, the valence band potential manifests a relatively high band offset of 0.64 eV, and this, combined with the heavier hole effective mass, yields minimal hole hybridization.

Considering the case of one exciton residing in the dimer and taking into account the Coulombic interaction between the electron-hole pair, since the hole wave-function is essentially not hybridized, the hole is in one of the dots and consequently the electron does not see a symmetric double QDs' potential anymore. The calculated two lowest energy levels wave-functions of the electron including the Coulombic interaction are presented in Fig.1h. The Coulombic interaction for the first electron level, localized around the hole, is greater than the second electronic state in the opposite dot, increasing the energy spacing between the bonding and anti-bonding states to 60 meV. One can see that the electron is localized in the dot which contains the hole as well. However, there is still significant tunneling-coupling observed for the electron wavefunction and a red shift is predicted. This is indeed confirmed experimentally in the emission and absorption spectrum, where only in the case of the fused dimer a red shift is observed compared to the monomer (603-607 nm in case of 1.4/2.1 nm core/shell QD), whereas, for the unfused organically linked dimer no red shift is seen (Fig. 1i and Supplementary Table 2). The control in the magnitude of the red-shift, for the monomer to fused dimer transition, for three different types of CQDs is depicted in Fig. 1j. In the the case of 1.2/2.1 nm core/shell CQDs the red shift is increased (13/14 meV calculated/experimental) due to the greater spill out of the electron wave-function to the shell because of the smaller core size. This is in contrast to the case of 1.9/4 nm core/shell CQD where the red shift is negligible (0.5/0 meV calculated/experimental) because of the localization of the electron wave-function in the larger core (Supplementary Fig.13, and Supplementary Table 2).

An additional signature for the coupling in fused dimers is observed in the absorption spectra at higher energies. Figure 1i shows broadening only upon fusion consistent with coupling forming multiple states in dimers. Furthermore, the spectra normalized at the band gap manifest a significantly stronger relative absorbance in high energies for the fused dimers compared with monomers and unfused dimers (see also Supplementary Fig.11). This is assigned to the wavefunctions modification in the fused system, which can be considered from a viewpoint of hybridization among the excited states.

**Structural characterization of fused CQD dimers**

We next consider further the nontrivial fusion stage and its consequences. Analysis by HAADF-STEM confirms that coupled dimer formation is indeed achieved based on fusion of the 1.9/4.0 nm core/shell QD monomers (Fig. 2). A continuous atomic lattice through the entire structure was formed upon fusing the two QDs shells (Fig. 2a). The core architecture in the coupled structure was maintained as demonstrated by the energy dispersive X-Ray spectroscopy (EDS) line scan measurement (Fig. 2b-c). A continuous distribution of cadmium (both in core and shell) and sulfur (only in shell) is identified along the line of the dimer axis. Along the same line, selective regions of the selenium (only in core) are clearly identified signifying the cores locations.

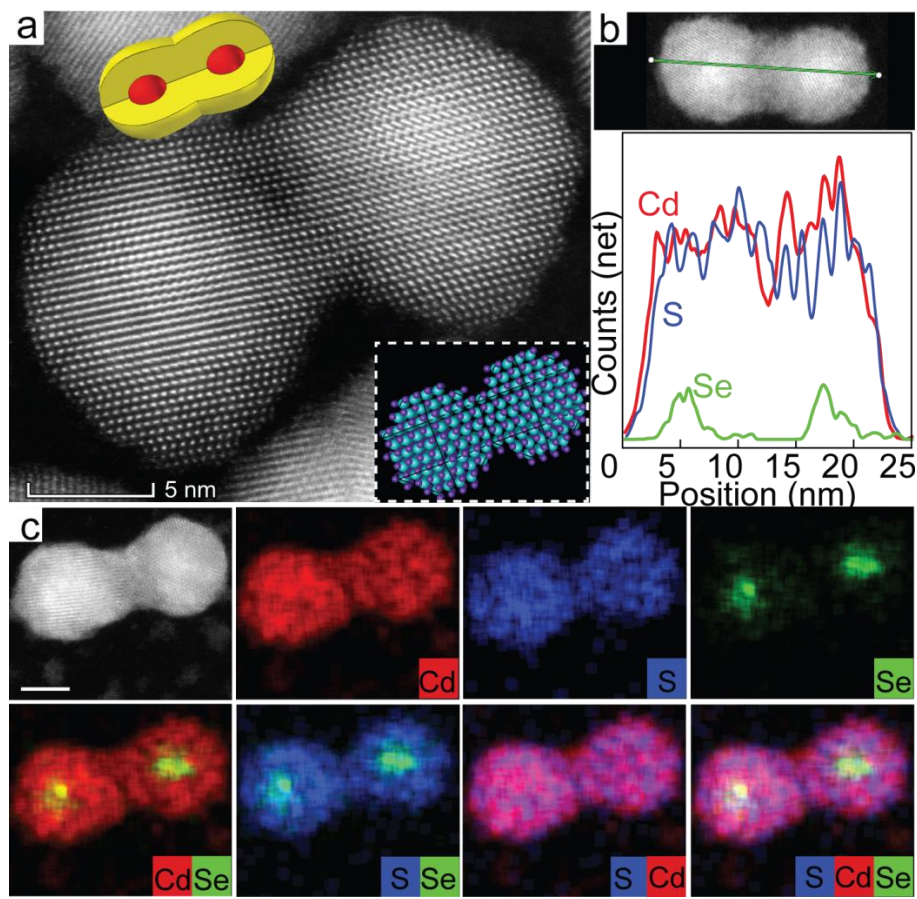

**Fig. 2**. **Dimer structure characterization and analysis.** HAADF-STEM image (a), EDS line scan data (b) and STEM-EDS (c) analysis of the coupled 1.9/4.0 nm CdSe/CdS molecules. Inset presents an atomic model. Scale bars 5 nm.

Underlying our fusion reaction strategy, is the process of oriented attachment - a crystal growth mechanism in which secondary mono-crystalline particles can be achieved through oriented and irreversible attachments of primary particles [24–30]. PbSe CQD dimers were prepared via oriented attachment in solution, but even under succinct control of ligands, concentration and reaction conditions, only ~30% dimer fraction, a rod-like colloidal quantum system, could be achieved along with monomers and higher order oligomers. In our template based strategy, high control over dimer formation was achieved by firstly forming a connection by molecular linkers. The molecular linkers however constrain the initial relative crystal orientations between the two monomers. With careful tuning of precursor and judicious choice of the fusion condition, we can foresee high potential for this method to serve as a general coupling strategy for other colloidal nanocrystal systems. Here, we studied this special case of "constrained oriented attachment", and Fig. 3 shows exemplary orientation relationships observed for coupled molecules formed from the 1.9/4.0 nm CQDs and their detailed analysis. Both homo-plane and hetero-plane (misorientation) attachment relationships are observed. Homo-plane attachment orientation occurs via contact between homonymous faces (10$\bar{1}$0) and (10$\bar{1}$0), (0002) and (0002), (10$\bar{1}$1) and (10$\bar{1}$1) (Fig. 3a-c, g, i, respectively), consistent with the CQD monomer crystal model built based on STEM analysis (Supplementary Fig.3). In such homo-plane attachment cases, both CQD monomers of a fused pair are projected under the same zone axis. This allows accurate identification of the fused faces at dimers orientated with its fusion axes normal to the projection zone axis (depicted in Fig.3). Hetero-plane attachment orientation is observed at fusion of heteronymous faces: (0002)||(10$\bar{1}$0) (Fig. 3e), (0002)||(10$\bar{1}$1), (10$\bar{1}$1)||(10$\bar{1}$0) (Supplementary Fig.14a,c).

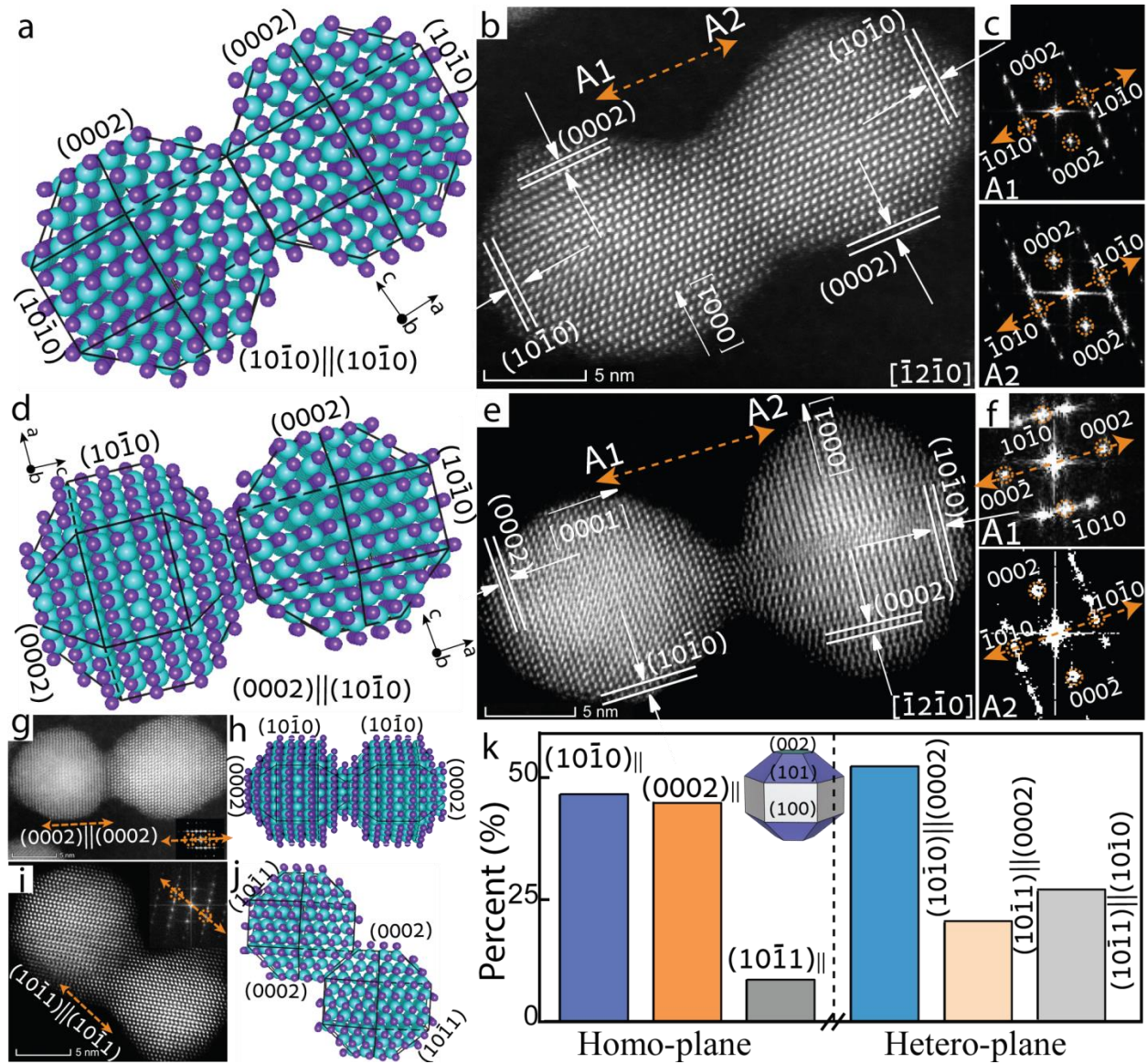

**Fig. 3**. **Fusion orientation relationships in CQD molecules.** Atomic structure model (a,d; cadmium atoms - brown, sulfur atoms - blue.), HAADF-STEM images (b,e), and FFT patterns (c,f) of the homo-

plane (a-c) and hetero-plane (d-f) attachment of coupled CQD molecules with orientation relationship of attachment on $(10\bar{1}0)\|(10\bar{1}0)$ and $(0002)\|(10\bar{1}1)$, respectively. The HAADF-STEM (g, i) and atomic structure model (h, j) of homo-plane attachment on (0002), and $(10\bar{1}1)$ facets, respectively. Dashed orange arrows indicate the CQD fusion/molecular axis in plane of the image normal to projection ZA $[\bar{1}2\bar{1}0]$. Note that for $(10\bar{1}0)\|(10\bar{1}0)$ homo-plane attachment, the homonymous $(10\bar{1}0)$ faces of A1 and A2 are parallel (c), while for the hetero-plane attachment the heteronymous faces are parallel $(0002)\|(10\bar{1}0)$ (e). (k) Distribution of observed homo- and hetero-plane attachment orientations on $(10\bar{1}0)$, (0002), and $(10\bar{1}1)$ faces. Here, the total amount of dimers for the statistic was 100. Inset shows the CQD model and faces.

The statistics of the orientation relationship within the CQD molecules is depicted in Fig. 3k (homonymous and heteronymous orientations are approximately equally abundant). The $(10\bar{1}0)\|(10\bar{1}0)$ and $(0002)\|(0002)$ face attachments are dominant, whereas the $(10\bar{1}1)\|(10\bar{1}1)$ attachment is much less common. This is consistent with an interplay between the relative reactivity, surface passivation and occurrence of the various faces on the monomer QDs. The (0002) facets, while in minority, manifest a Cd rich termination with 3 dangling bonds per atom, that can easily react with thiol linkers [31]. Both $(10\bar{1}0)$ and $(10\bar{1}1)$ facets are plentiful but better passivated [32,33]. However, linking to the $(10\bar{1}1)$ facet is sterically hindered. The hetero-plane attachment statistics is also consistent with these considerations.

The generality of our formation strategy is well manifested also for the other CQD homodimers. Analytical and structural STEM analysis for the 1.4/2.1 nm core/shell CQDs are shown in Fig. 4 and Supplementary Fig.15-16. Fusion conditions need to be tuned as smaller core/shell CQDs are more reactive. Therefore, shorter etching times were used, and also a shorter fusion process (~10h).

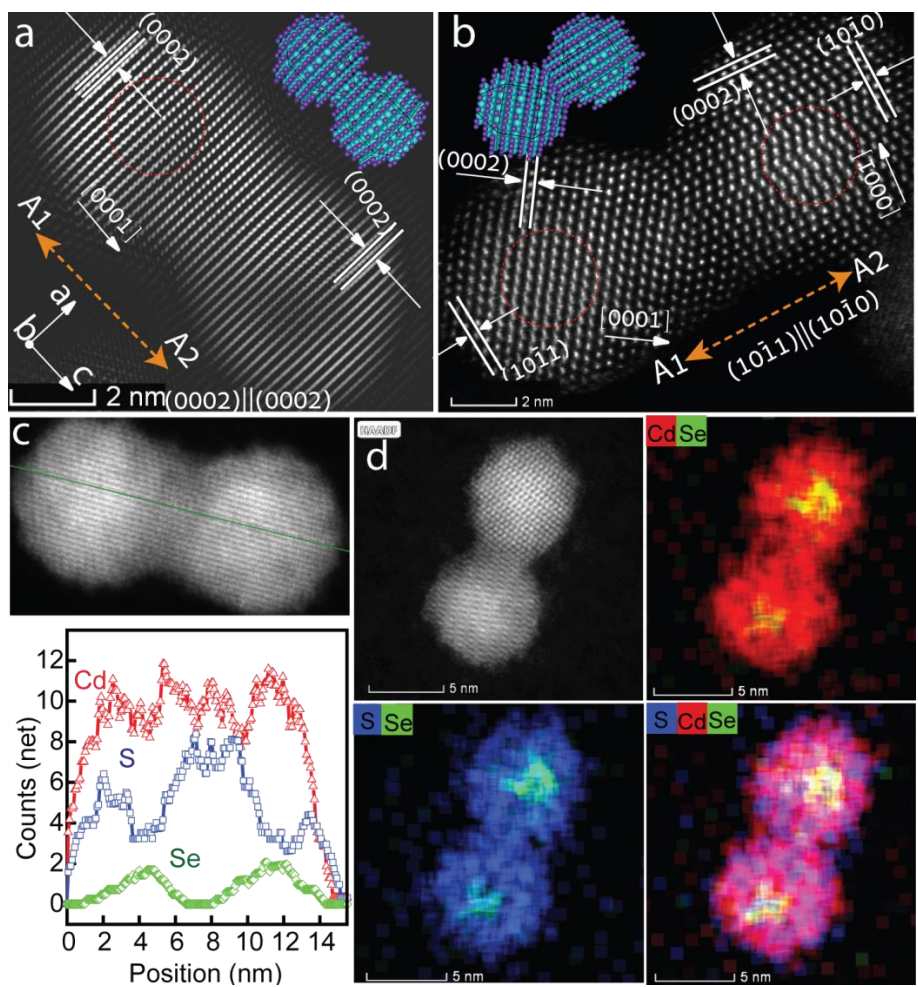

**Fig. 4. Characterization of 1.4/2.1 nm CQDs molecule.** Fourier filtered HAADF-STEM images of the coupled CdSe/CdS molecules with homo-plane attachment of (0002)‖(0002) (a) and hetero-plane attachment of (10$\bar{1}$0)‖(10$\bar{1}$1) faces (b). (c) EDS line scan data and (d) STEM-EDS analysis.

**Photophysical characteristics of CQDs molecule**

Further manifestations of coupling are observed in the fluorescence properties studied both in ensemble and as single particles, comparing monomers with unfused and fused dimers. For the 1.4/2.1 nm CQDs the fluorescence decays vary, with monomers showing a nearly single exponential lifetime of 25ns whereas fused dimers show nearly bi-exponential decay with a fast component of 5ns (Supplementary Table 3). Unfused dimers show an intermediate behavior. Molecules of the larger 1.9/4.0 nm CQDs exhibit significantly smaller changes in lifetimes upon fusion, supporting the role of coupling in the variations observed for the 1.4/2.1nm CQDs (Supplementary Fig.17).

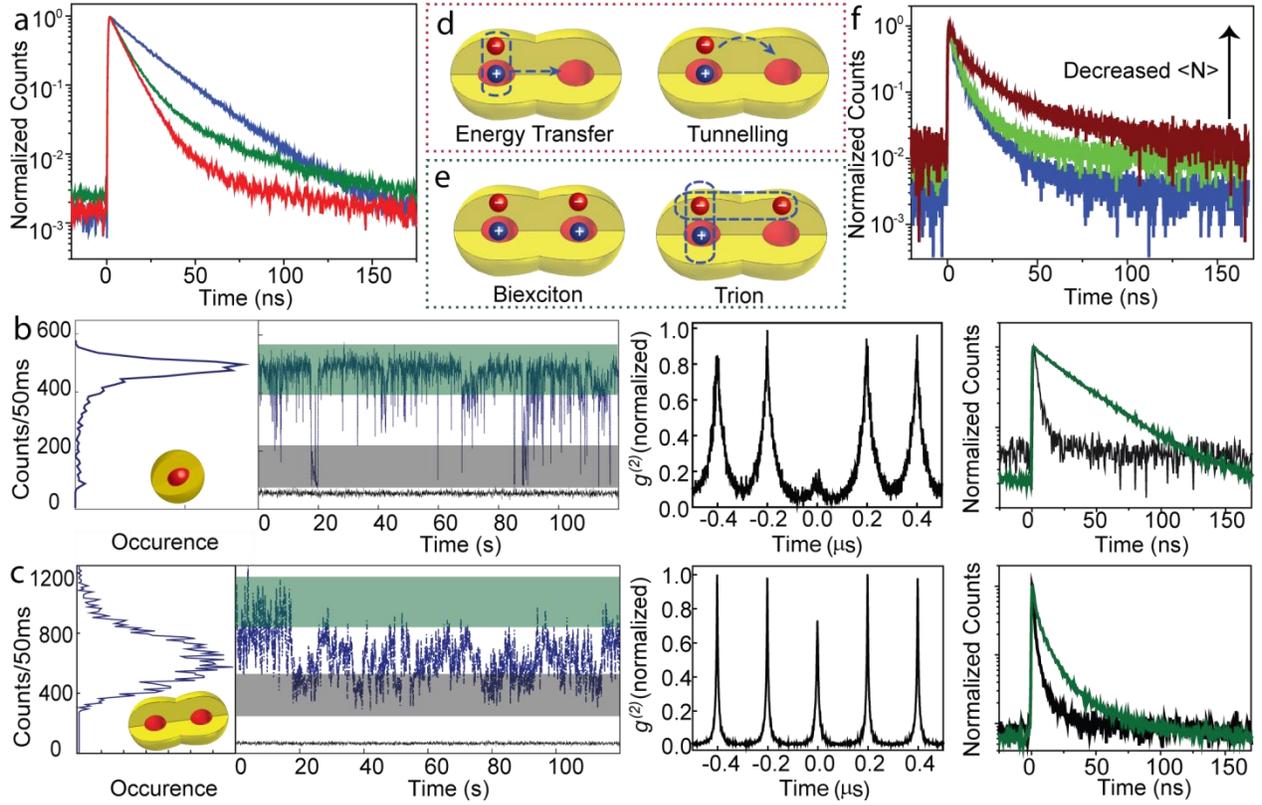

**Fig. 5**. **Coupling effects through fluorescence measurements for 1.4/2.1 nm CQDs molecules.** (a) Ensemble photoluminescence lifetime decay for monomer (blue line), unfused dimer (green line) and fused dimer (red line). Time-tagged, time resolved data for single (b) CQD monomer and (c) fused dimer. Shown are, (from left to right) photoluminescence intensity time trace, second order photon correlation and lifetime for the single particle respectively (the green and dark grey lifetime curves were generated from data shaded in the same color in the corresponding time traces). The excitation fluence (<N>) for monomer and coupled dimer in the represented figures were 0.05 and 0.08, respectively. (d) Possible coupling mechanisms for shortening of lifetime in molecules. (e) Multicarrier configurations. (f) Pump fluence dependency of lifetime for an individual fused dimer. <N> values are 0.03 (brown trace), to 0.09 (green trace) and 0.18 (blue trace).

More detailed information is garnered from single nanoparticle fluorescence measurements on the 1.4/2.1 nm CQDs ('Methods' section for experimental details). The fluorescence from single monomers (Fig. 5b) exhibited a typical on-off bimodal distribution with a monoexponential fluorescence decay of the on-state [34], and strong photon antibunching [35]. In comparison, individual fused dimers (Fig. 5c) under similar excitation conditions show more intense fluorescence consistent with their two-fold larger absorption cross section. However, flickering of the fluorescence with a distribution of intensities is observed rather than distinct on-off states, accompanied by a significantly lower antibunching contrast (~0.75). The lifetime measurements indicate systematic shortening of the average single particle lifetimes from the monomers through the unfused to the fused dimers, in line with the ensemble measurements (Supplementary Fig.18 for representative traces and statistics of the lifetimes). Analyzing the lifetimes of the high intensity occurrences (green shaded regions in Fig. 5b,c), yields a significantly shortened average lifetime of 5ns for the single fused dimer compared with the monomer (29ns). It is noteworthy that within the fused dimers sample, we have detected ~15% of particles that have similar fluorescence characteristics as the CQD monomer sample, in line with their fraction from TEM analysis (Supplementary Fig.19-20). This establishes that the fusion procedure in its entirety did not change the core/shell CQDs.

We consider two possible mechanisms related to coupling, both leading to shortening the lifetime in dimers. First, resonance energy transfer between the two dots (Fig. 5d), a mechanism nearly equally active for unfused and fused dimers. Second, tunneling of the electron to the other dot (Fig. 5d) as already illustrated in Fig. 1. Tunneling in unfused dimers occurs by collisional electron transfer and is strongly dependent on the linker [36], while in the fused dimer the potential barrier for tunneling is modified and reduced substantially. Both mechanisms will be enhanced for the smaller CQD dimers, but tunneling is more strongly dependent on the size/distances. Indeed, the large CQD molecules, where tunneling probability is negligible according to our calculations (Fig. 1), show smaller changes in lifetimes with shortening in the case of dimers compared to monomers and little differences before and after fusion (Fig. 5c, Supplementary Fig.17, 21) the notable lifetime shortening that is seen for the small CQD molecules upon fusion is indicative of the enhanced contribution of the tunneling mechanism in this case.

Next, we consider the photon statistics, which in CQDs are strongly influenced by multicarrier effects. An increment in the $g^2(0)$ value was found in the case of dimers in general with the possibilities of either two emission centers in the excitation spot or intrinsic properties of coupled systems. Specifically, in the fused dimers, the particles absorb the light as one unit, and at low excitation regime (<N>~0.1) the possibility of emission from one of the centers is rational statistically. The full understanding of the $g^2(0)$ value requires more rigorous experimental studies while an interesting multicarrier configuration can be realized in these fused CQD dimers. In dimer CQD molecules, a new type of biexcitons can occur, with each exciton occupying a different core (Fig. 5e). The large increase in the value of of [$g^2 = \frac{Area_{(0ns)}}{Area_{(200\,ns)}} = \frac{QY_{(BX)}}{QY_{(X)}}$ at low excitation power] observed for the dimers versus monomers can be explained by this new type of biexciton for which the non-radiative Auger decay will be strongly suppressed increasing the biexciton quantum yield (QY) (Supplementary Fig.22). Moreover, the single particle exciton QY will decrease for dimers on account of the tunneling of the electrons reducing the electron-hole overlap.

An additional difference relates to the fluorescence flickering in the dimers rather than distinct on-off fluorescence of the monomers (Fig. 5c, Supplementary Fig.23). This indicates presence of multitude emitting configurations for dimers. Indeed, the lifetime traces for the high/low intensity regions in the dimer are not single exponential. The low intensity region is above background and not off and the lifetime has a ~5ns component. All this indicates to trion formation (positive or negative). In small CQD monomers, the trion states are strongly quenched by the Auger decay yielding an off state behavior. In dimers, which have large volume and the excess carrier may occupy the second dot region forming a new type of trion (Fig. 5e), the Auger rate is suppressed, and the trion can become emissive. Such an effect was reported for large CdSe/CdS core/shell manifesting gray state emission and is also observed by us for the large CQDs [37]. The multitude possibilities for emissive trion formation can explain the larger distribution of observed fluorescence intensities and the lifetime behavior for the dimers. To further address this point, an excitation intensity dependence was performed varying the average exciton occupancy <N> from 0.03 to 0.18 (Fig. 5f, Supplementary Fig.24). The lifetime decreases upon increasing the laser power, indicating the increasing contribution of trion formation consistent with this description [34,38].

This study introduces a robust route to the realization of coupled colloidal quantum dot molecules with high purity. The construction strategy for the CQD molecules utilizes well-controlled CdSe/CdS core/shell nanocrystal artificial atom building blocks constructing homodimers fused via "constrained attachment" of the nanocrystal facets. The synthesized novel CQD dimer structures, emitting in the visible range, exhibit electronic coupling at room temperature signified by the red-shift and broadening of the band-edge transition observed both in absorption and in photoluminescence. This is attributed to the quantum coupling and hybridization of the monomer wavefunctions within the CQD molecules, as supported by quantum mechanical calculations. Broadening and modification of the absorption transitions at higher energies are also related to the coupling effects within the CQD dimer molecules. Furthermore, the photophysical properties of the CQD molecules at single nanoparticle level exhibit lifetime shortening, fluorescence flickering, and an increase in the g2 value for the photon antibunching compared to monomers. These characteristics indicate the introduction of additional recombination pathways and the rich possibilities for multiexciton configurations in the artificial CQD molecule compared with the monomers, also related to coupling within the system.

Formation of coupled CQD homodimer molecules sets the stage for *Nanocrystal Chemistry*. Considering the rich selection of size and composition controlled CQDs emphasizes the analogy of these artificial atoms to atoms of the periodic table discovered by Mendeleev 150 years ago. As a future outcome, we foresee the formation of a diverse variety of coupled CQD molecules with prodigious promise for their utilization in numerous optoelectronic, sensing and quantum technologies applications.

# Methods

### Reagents

Oleylamine (OAm, 70%), oleic acid (OA, 99%), 1-octadecene (ODE, 90%), cadmium oxide (CdO, ≥99.99%), selenium (Se, 99.99%), trioctylphosphine oxide (TOPO, 99%), 1-octanethiol (≥98.5%), pentaerythritol tetrakis(3-mercapto-propionate) (95%), ammonia aqueous (28.5%), tetrahydrofuran (THF, ≥99.9%), N-methylformamide (NMF, 99%), toluene (99.8%), hydrofluoric acid (HF, 48%), polyvinylpyrrolidone (PVP, 10k), ethanol (99%), tetraethyl orthosilicate (TEOS, 98%), hexane (95%) and (3-Mercaptopropyl) trimethoxysilane (MPTMS, 95%) were obtained from Sigma Aldrich. Trioctylphosphine (TOP, 97%) was purchased from Strem Chemicals. Octadecylphosphonic acid (ODPA, >99% was purchased from PCI synthesis. All the reagents were used as received without further purification.

### CdSe core growth

Briefly, 60 mg CdO, 280 mg ODPA and 3 g TOPO were added to a 50 mL flask. The mixture was heated to 150 °C and degassed under vacuum for 1 hour. Under argon flow, the reaction mixture was heated to 320 °C to form a colourless clear solution. After adding 1.0 mL TOP to the solution, the temperature was brought up to 350 °C. At this point Se/TOP solution (60 mg Se in 0.5 mL TOP) was swiftly injected into the flask. The reaction was kept at 350 °C for suitable time and then stopped by removal of the heating mantle. The resulting CdSe particles were precipitated with acetone and redispersed in 3 mL hexane for use as a stock solution.

### CdSe/CdS core-shell colloidal quantum dots synthesis

For the shell growth reaction, a hexane solution containing 200 nmol of CdSe colloidal quantum dots (CQDs) mixed with ODE (6 mL) and OAm (6 mL). The reaction solution was degassed under vacuum at room temperature for 30 min and at 90 °C for additional 30 min to completely remove the hexane, water and oxygen inside the reaction solution. Then the reaction solution was heated up to 310 °C under argon flow and magnetic stirring. During the heating, when the temperature reached 240 °C, a desired amount of cadmium (II) oleate (Cd-oleate, diluted in 6 mL ODE) and octanethiol (1.2 equivalent amounts to Cd-oleate, diluted in 6 mL ODE) was injected dropwise into the growth solution at a rate of 3 mL/h using a syringe pump. Upon precursor infusion, 2 mL oleic acid was quickly injected and the solution was further annealed at 310 °C for 30 min. The resulting CdSe/CdS core/shell CQDs were precipitated by adding ethanol, and then redispersed in hexane. The particles were further purified by additional two precipitation-redispersion cycles and finally suspended in ~2 ml hexane.

**Silica nanoparticles (NPs) synthesis**

120 µL MPTMS precursor was mixed with 30 mL ammonia aqueous solution (1%) under strong stirring. After stirring for 1 min, the solution was stored overnight. The $SiO_2$ NPs were collected by centrifugation and dispersed in 10 mL of ethanol. Then the $SiO_2$ solution was mixed with PVP solvent (0.02 g/mL) for 30min. Finally, the nanoparticles were stored after the cleaning by centrifugation.

**The synthesis of CdSe/CdS@$SiO_2$**

1mL of $SiO_2$ NPs (0.0079 g/mL) was mixed with 0.5 nmol CdSe/CdS NPs using vortex for 5 min. Then 5 mL of ethanol was added into the vails to precipitate and remove the unattached NPs. After three washing cycles the final $SiO_2$@CdSe/CdS NPs were redispersed in 5 mL of ethanol.

**The synthesis of $SiO_2$@CdSe/CdS@$SiO_2$**

The CdSe/CdS@$SiO_2$ was dispersed in 5 mL of ethanol. Then 330 µL of ammonia solvent (28.5% wt %) was added into the solution with stirring for 5 min. Thereafter, 50 µL of TEOS was added dropwise into the solution. After stirring for 10 h, the resulting solvent was centrifuged (6000 rpm, 5 min) and redispersed in 5 mL of THF.

**The synthesis of Dimer-CdSe/CdS@$SiO_2$**

A tetrathiol linker ,pentaerythritol-tetrakis (3-mercapto-propionate), (200 µL) was added to the CdSe/CdS@$SiO_2$ solution. Then 0.6 nmol of CdSe/CdS CQDs were added and the solution was stirred in an oil bath at 60 °C overnight. Samples were cleaned by centrifugation (6000 rpm,5 min) and redispersed with 10 mL of THF for storage.

**The release of Dimer-CdSe/CdS**

1 mL of Dimer-CdSe/CdS@$SiO_2$ CQDs was centrifuged (5000 rpm, 5 min), and later mixed with 2 mL of mixed HF/NMF solvent (10%) under stirring for 10 h. Upon etching, the color of the samples changed to light yellow, which indicates on the removal of the $SiO_2$. Thereafter, the samples were precipitated by centrifugation (6000 rpm, 10 min) and washed twice. Finally, the samples were redispersed in 2 mL of ethanol.

**The synthesis of fused Dimer-CdSe/CdS CQDs**

Dimer-CdSe/CdS CQDs (in 2 mL of ethanol) were mixed with 2mL of ODE, 100 μL of Cd-oleate (0.2 M), and 50 μL of OAm. The reaction solution was degassed under vacuum at room temperature for 30 min and again at 90 °C for additional 30 min. Later, the reaction mixture was heated to 180 °C for 20 h under argon flow. The resulting fused particles were precipitated with ethanol and redispersed in 2 mL toluene for use as a stock solution.

**Optical and structural characterization**

Absorption spectra were measured using a Jasco V-570 UV-Vis-NIR spectrophotometer. Fluorescence spectra and ensemble lifetimes were measured with fluorescence spectrophotometer (Edinburgh instruments, FL920). Transmission electron microscopy (TEM) was performed using a Tecnai $G^2$ Spirit Twin T12 microscope (Thermo Fisher Scientific) operated at 120 kV. High resolution TEM (HRTEM) measurements were done using Tecnai F20 $G^2$ microscope (Thermo Fisher Scientific) with an accelerating voltage of 200 kV. High resolution scanning-transmission electron microscopy (STEM) imaging and elemental mapping was done with Themis Z aberration corrected STEM (Thermo Fisher Scientific) operated at 300 kV and equipped with high angular annular dark field detector (HAADF) for STEM and Super-X energy dispersive X-Ray spectroscopy (EDS) detector for high collection efficiency elemental analysis. CQDs atomic structure model were built by the VESTA software. Scanning electron microscopy imaging (SEM) was done with HR SEM Sirion (Thermo Fisher Scientific) operated at 5 kV.

**Single particle optical measurements**

Single particle measurements were performed with an inverted microscope (Nikon Eclipse-Ti) in epi-luminescence configuration. Dilute solution of quantum dots in 2wt% poly(methyl-methacrylate) were spin coated on glass coverslips (no.1, precleaned and thermally annealed) leading to minimum separation of 4-5μm between the dots as confirmed by wide field fluorescence microscopy. The excitation light from a pulsed diode laser (EPL375; Edinburgh Instruments) at a repetition rate of 5MHz was focused onto single particles through an oil immersion objective (100X; 1.4 NA) which was also used for collecting the emission. The emission light was passed through a dichroic mirror (T387lp) and additional longpass filter (580LP) before focusing onto two Avalanche Photodiodes (Perkin-Elmer; SPCM-AQRH-14) in a Hanbury-Brown-Twiss geometry. Photon statistics of the signal from the detector were performed using multichannel Time Tagger 20 (Swabian Instruments). Time traces, fluorescence lifetime and second order photon correlation were extracted from the time-tagged time-resolved data by using home written MATLAB code.

**Data availability**

The authors declare that all the important data to support the findings in this paper are available within the main text or in the supplementary information. Extra data are available from the corresponding author upon reasonable request.

# Acknowledgments

The research leading to these results has received financial support from the European Research Council (ERC) under the European Union's Horizon 2020 research and innovation programme (grant agreement No [741767]). J.B.C and S.K acknowledge the support from the Planning and Budgeting Committee of the higher board of education in Israel through a fellowship. U.B. thanks the Alfred & Erica Larisch memorial chair. Y.E.P. acknowledges support by the Ministry of Science and Technology & the National Foundation for Applied and Engineering Sciences, Israel.

# Author information

**Affiliations**

Institute of Chemistry, The Hebrew University of Jerusalem, Jerusalem 91904, Israel.

The Center for Nanoscience and Nanotechnology, The Hebrew University of Jerusalem, Jerusalem 91904, Israel.

**Author contributions**

U.B, oversaw and managed the research. J.B.C, Y.E.P, S.K, M.O and U.B. designed the experiments. J.B.C. performed the synthesis and materials characterization. S.K and Y.E.P carried out single molecules experiments. Y.E.P. and D.S performed the calculation. S.R. and I.P. assisted with HR-STEM measurements and structural analysis. J.B.C, S.K, Y.E.P, N.W, and U.B. wrote the manuscript with help from the other authors.

**Competing Interest**

The authors declare no competing interests.

**Corresponding author**

Correspondence and requests for materials should be addressed to U.B. (uri.banin@mail.huji.ac.il)

**Additional information**

Reprints and permissions information is available at www.nature.com/reprints. Readers are welcome to comment on the online version of this article at www.nature.com/nature.

## Supplementary information

Supplementary Information is linked to the online version of the paper at www.nature.com/nature.

## Rights and permissions

Supplementary Information for

# Colloidal Quantum Dot Molecules Manifesting Quantum Coupling at Room Temperature

Jiabin Cui[†], Yossef E. Panfil[†], Somnath Koley[†], Doaa Shamalia, Nir Waiskopf, Sergei Remennik, Inna Popov, Meirav Oded & Uri Banin*

[†] These authors contributed equally to this work.

* Corresponding author. Email: uri.banin@mail.huji.ac.il (U.B.).

**Contents:**

Supplementary Discussion

Supplementary Figure 1 to 24

Supplementary Tables 1 to 3

Supplementary References

**Supplementary Discussion**

*Electronic structure calculation methodology:* We have calculated the energy levels and the electron and hole wave-functions of the CdSe/CdS monomer and dimers using the multiphysics mode of COMSOL.[1,2] Electron and hole states are calculated with a 3D single-band effective mass Hamiltonian. Interacting electron and hole states are obtained by iterative solution of the Schrödinger-Poisson equations, within a self-consistent Hartree procedure, taking into account the dielectric mismatch between the surroundings and the CQD. The dimensions we used in the calculation are based on experimental data. Material-dependent parameters such as effective masses $m^*_{e,h}(r)$ dielectric constants $\varepsilon(r)$, and conduction and valence band profiles ($V^{e,h}_{confinement}$) used in this calculation are summarized in Supplementary Table 1. The entire computational space extends further from the QD boundary allowing for electron and hole wavefunctions to extend outside of the QD boundaries and decay into free space.

We start the simulation by computing the non-interacting electron and hole states by solving the Schrödinger equation:

$$\left(-\frac{\hbar^2}{2}\nabla\left(\frac{1}{m^*_{e,h}(r)}\nabla\right) + V^{e,h}_{confinement}(r)\right)\Psi^n_{e,h} = E^n_{e,h}\Psi^n_{e,h}(r) \quad (1)$$

We use von Neumann boundary-condition at the inner (between core-shell) and outer boundaries of the QD in order to impose the Ben-Daniel-Duke condition. At the edge of the computational domain (around 100 nm away from the QD) we set the Dirichlet boundary-condition by setting the wavefunction to zero. After normalization of the wavefunctions, the Poisson equation is solved to derive the hole or electron Coulombic potential generated from the other particle $\phi_{e,h}(r)$.

$$\nabla(\varepsilon_0 \cdot \varepsilon(r)\nabla\phi_{e,h}(r)) = q_e <\Psi_{e,h}(r)|\Psi_{e,h}(r)> \quad (2)$$

With these potentials, Schrödinger equations are solved again for the electron and the hole with all of the contributions to the potentials:

$$\left(-\frac{\hbar^2}{2}\nabla\left(\frac{1}{m^*_e(r)}\nabla\right) + V^e_{confinement}(r) + q_e \cdot \phi_h(r)\right)\Psi^n_e(r) = E^n_e\Psi^n_e(r) \quad (3)$$

$$\left(-\frac{\hbar^2}{2}\nabla\left(\frac{1}{m^*_h(r)}\nabla\right) + V^h_{confinement}(r) + q_e \cdot \phi_e(r)\right)\Psi^n_h(r) = E^n_h\Psi^n_h(r) \quad (4)$$

This process is repeated iteratively until the electron and hole energies converge. In most cases, three iterations are sufficient to obtain a convergence. The emission energy was calculated by:

$$E_{emission} = E_g + E^i_e + E^i_h - E_{coulomb} \quad (5)$$

$$E_{coulomb} = \frac{E_e^i - E_e^f + E_h^i - E_h^f}{2} \quad (6)$$

Where $E_g$ (1.76 eV) [3] is the energy gap of CdSe. $E_{emission}$ is calculated in a way which avoids the consideration of the coulomb potential twice (both for the electron and both for the hole).

**Supplementary Table 1.** Material parameters used in the simulations.

| | CdSe | CdS | Environment | Units | Ref. |
|---|---|---|---|---|---|
| $V_{confinment}^e$ | 0 | 0.1 | 5.1 | [eV] | (3) |
| $V_{confinment}^h$ | 0 | -0.64 | -5.64 | [eV] | (3) |
| $m_e^*$ | 0.112 | 0.21 | 1 | $m_0$ | (4) |
| $m_{h\perp}^*$ | 0.48 | 0.376 | 1 | $m_0$ | (4) |
| $m_{h\parallel}^*$ | 1.19 | 0.746 | 1 | $m_0$ | (4) |
| $\varepsilon_\perp$ | 9.29 | 8.28 | 1 | - | (4) |
| $\varepsilon_\parallel$ | 10.16 | 8.73 | 1 | - | (4) |

*Consideration of Biexciton quantum yield (QY) to exciton QY ratio in fused dimers:* To consider the obtained $g2(0)$ value as the biexciton quantum yield, we kept the excitation power significantly low so that the average number of excitons produced per pulse (<N>) always remained below 0.2 [5].

The average number of excitons produced per pulse is given by the formula $<N> = \frac{\sigma J}{E_{Photon} \cdot A}$ where σ is the absorption cross section of the particle at the excitation wavelength (375 nm in our case), *J* and $E_{Photon}$ are the pulse energy and photon energy of the excitation light, respectively, and *A* is the spot area. The absorption cross-section was calculated according to the procedure reported elsewhere. [6] The value of the absorption cross-section for the 1.4/2.1 nm CQDs and the corresponding fused dimer at 375nm was found to be $2.3 \times 10^{-15}$ cm$^2$ and $5.7 \times 10^{-15}$ cm$^2$, respectively.

*Excitation power dependence study on single fused dimer:* The flickering nature of the fluorescence in single fused dimers associated with a tri-exponential lifetime decay (comprising of ~2 ns, ~5 ns and ~25 ns components) was observed for the 1.4/2.1 Coupled CQDs. The fluorescence flickering nature was present upon changing the laser power and observed even at the lowest measurable excitation. The intensity traces followed a single distribution of intensities (much narrower than the high excitation) without reaching an off state (Supplementary Figure 24). Upon excitation of the particle by a short pulse laser, a charge state of the exciton can be created along with the neutral exciton which can contribute to the appearance of short decay components [7, 8].

For the 1.4/2.1 nm monomer CQDs the contribution of the short fluorescence decay component was found to be negligible at similar excitation condition, whereas the decay for dimers comprised of these CQDs was dominated by the short components. With decrease in the excitation intensity, we observed an enhancement in the contribution of the ~25 ns component on account of the ~2 ns and ~5 ns components (Fig. 5f in the main manuscript). This clearly indicates on the enhanced photo-charging effect in case of fused dimers. Associated with the flickering of the intensity, we observed a lifetime distribution through the intensity traces (lower intensity traces have shorter lifetime as shown in figure 5c). Upon the decrease in photo-charging not only the average lifetime increases (from 4.5 ns at <N>~0.18 to 7.2 ns at <N>~0.03), but also the lifetime distribution (commonly referred as lifetime flickering) narrows down.

**Formation of coupled CQDs molecule dimers**

We utilized silica nanoparticles as a template [9] for the fabrication of coupled CQD molecules. The detailed route-line is depicted in Supplementary Figure 1. This was performed by the following steps:

1. Fabrication of $SiO_2$ nanoparticles, coated by MPTMS. This kind of $SiO_2$ particle presents thiol groups on its outer surface, which are later used for the binding of the CQDs.
2. Core/shell CQDs binding to the $SiO_2$ particle surface: mixing a solution of the chosen core/shell CQDs with the $SiO_2$ nanoparticles allows their binding to the available thiol sites on the silica surface.
3. Growth of a secondary thin layer of $SiO_2$ on the QDs@ $SiO_2$ for masking: In this manner the CQDs are immobilized and cannot rotate or reorient while only a top hemisphere is remained exposed for further chemical functionalization of the CQDs.
4. Selective surface decoration of the CQDs by linker groups: Chemical grafting of a functional structure based on thiol group as linkers is then applied which selectively reacts only with the exposed NC hemisphere. Here, a tetrathiol ligand was added as a linker by a ligand exchange reaction on the exposed CQDs surface (for example oleyamine).
5. Forming the dimer geometry on the silica surface: Addition of a solution with the second CQDs allows their conjugation, yielding a controlled formation of a dimer structure through the binding linker.
6. Dimer release: The dimers are released from the silica surface and separated. Here, this can be achieved by selective $SiO_2$ etching using an HF/NMF solution.
7. Fusion to form the coupled CQD molecule: The dimers are fused by addition of a suitable precursor and heated to form a continuous link between the two shell regions of the pre-made dimers.

Additional optional purification (size selective separation) of the dimers from free monomer and higher linked oligomers is possible in between steps 6 and 7, or after the fusion step.

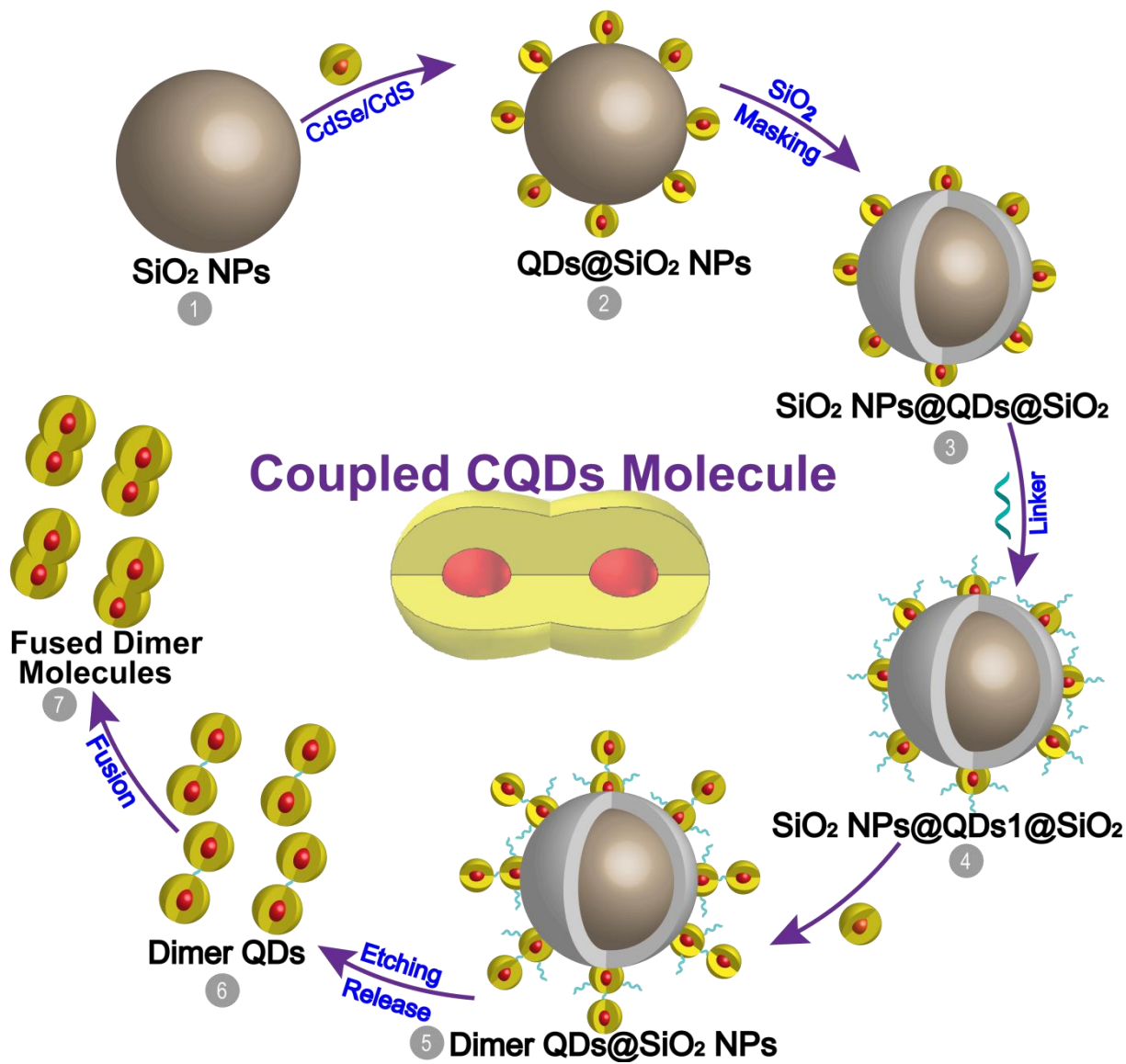

**Supplementary Figure 1. Scheme of the coupled CQDs molecule fabrication steps.** Step 1: SiO$_2$ nanoparticles synthesis. Step 2: core/shell CQDs binding to the SiO$_2$ particle surface. Step 3: masking by a secondary thin layer of SiO$_2$ growth on the QDs@SiO$_2$. Step 4: selective surface decoration of the CQDs by linker groups. Step 5: dimer geometry formation on the silica surface. Step 6: dimer release. Step 7: fusion to form coupled CQD molecules.

**CdSe@CdS Core/Shell NCs synthesis and characterization:**

Supplementary Figure 2 shows electron microscopy images and the absorption and emission spectra for the different nanocrystal building blocks which were used in this work. For example, Supplementary Figure 2a-c, show CdSe cores (1.9 nm, radius) which were synthesized by the hot-injection method. The absorption (first exciton peak at 581 nm) and photoluminescence (PL) spectra (peak at 597 nm) were measured after purification. The CdSe@CdS CQDs (1.9/4.0 nm) were fabricated by CdS shell growth using the injection of Cd-oleate and octanethiol precursors. The PL spectrum (637 nm) after shell growth, shows a red-shift compared with the cores emission. This, along with the enhancement of the fluorescence quantum yield indicates the successful fabrication of quasi-type II QDs.

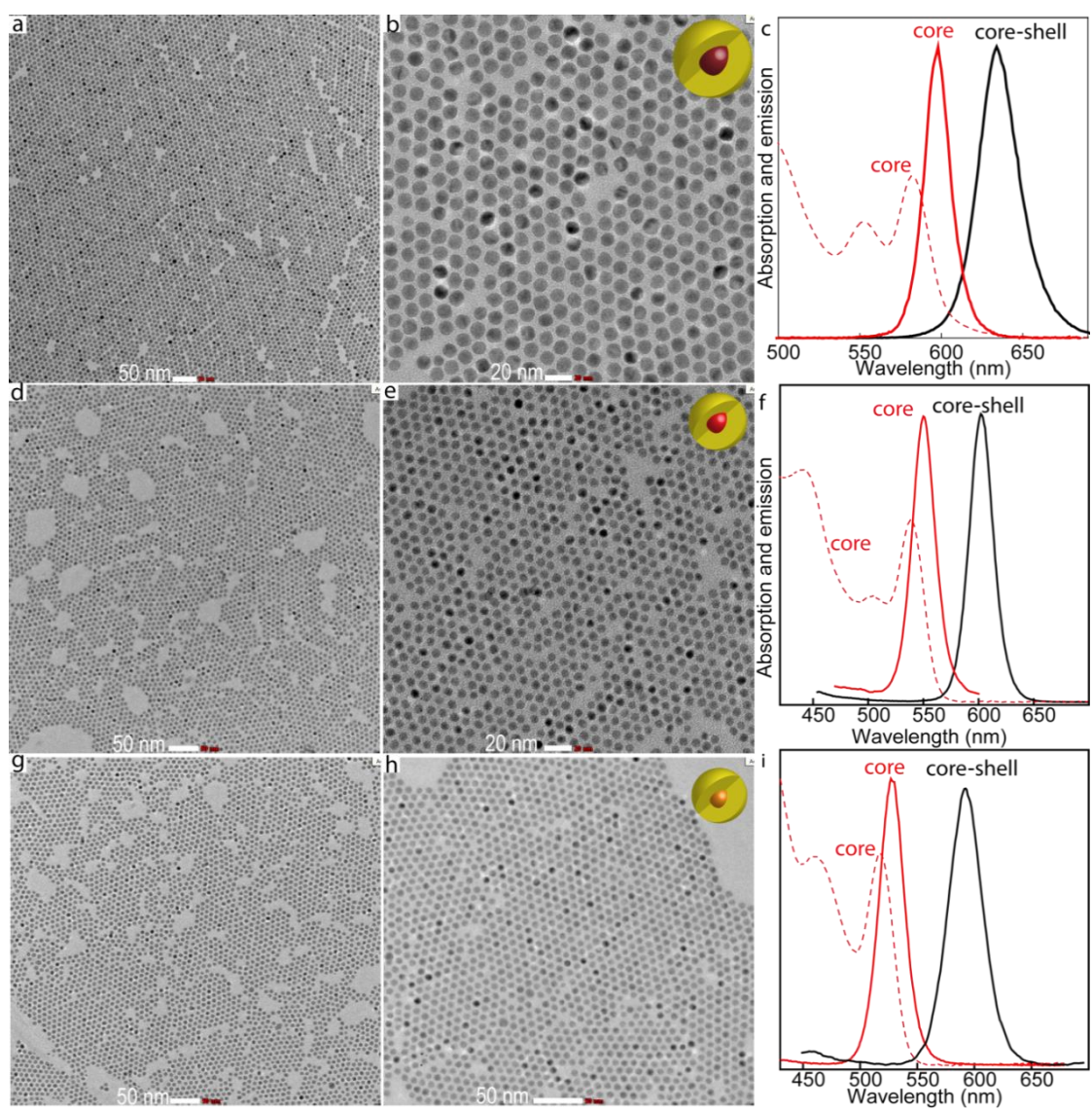

**Supplementary Figure 2. Quantum dots characterization.** TEM micrographs, absorption and photoluminescence spectra of different sizes of CdSe/CdS core-shell CQDs: (a-c) 1.9/4.0 nm, (d-f) 1.4/2.1 nm, and (g-i) 1.2/2.1 nm.

**Structural characterization**

The core/shell structure of the monomer is directly identified by atomic resolution STEM-HAADF imaging followed by fast Fourier transform (FFT) analysis (Supplementary Figure 3). Supplementary Figure 3a, b shows raw and Fourier filtered images of CQD monomers, respectively, viewed under [$\bar{1}2\bar{1}0$] zone axes (ZA) of their wurtzite structure. At this orientation major low index atomic planes (0002), (10$\bar{1}$0), (10$\bar{1}$1) are clearly observed thus providing identification of boundary faces of a monomer crystal (labeled on inset in (a)). Atomic structure of CQDs at [0001] ZA with clearly visible perfect hexagonal motifs is depicted in Supplementary Figure 3g-h. The hexagonal close-packed (*hcp*) wurtzite structure of CQDs was further evidenced by the selected area electron diffraction (SAED) and X-ray powder diffraction (XRD) measurement (Supplementary Figure 3k-l).

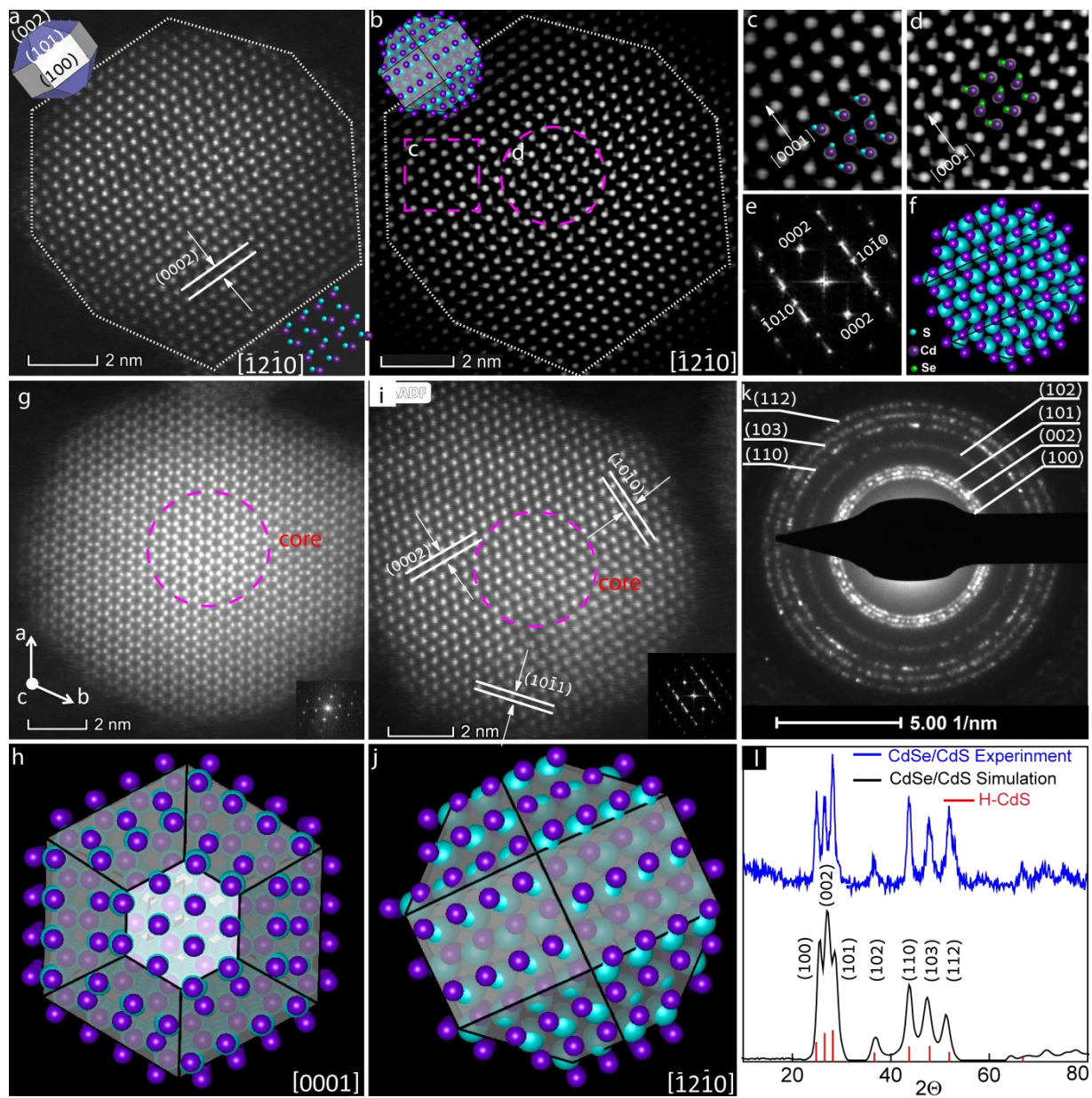

**Supplementary Figure 3. Structural characterization of the coupled CQDs.** Raw (a) and Fourier filtered (b) HAADF-STEM images of 1.9/4.0 nm CdSe/CdS CQD monomer viewed under [$\bar{1}2\bar{1}0$] zone axis (ZA). Inset in (a) is a cartoon model built with VESTA software [10] with bounding faces indexed based on the STEM data. Magnified images of edge (shell) (c) and central (core) parts (d) of the CQD shown in (b). Sulfur, Selenium and Cadmium atoms are marked in blue, green and purple, respectively. Coherent growth of the shell lattice is identified. (e) and (f) are FFT and atomic structure model of (a), respectively. HAADF-STEM image of CdSe/CdS CQD under ZA [0001] (g) and atomic structure reconstruction imaging calculated for the same orientation (h). (i) High resolution HAADF-STEM image and atomic structure model (j) of CdSe/CdS CQD viewed under ZA [$\bar{1}2\bar{1}0$]. The core regions are marked with pink circles in (g) and (i). FFT patterns are inserted in (g), and (i). SAED (k) and XRD pattern acquired at large ensembles of CdSe/CdS CQDs (blue curve – experimental XRD data, red bars - theoretical positions for diffraction peaks of *hcp*) CdS (JCPDF 04-001-6853), black curve - integrated intensity of SAED (k)).

**Step I - Silica nanoparticles characterization**

The SiO$_2$ nanoparticles (step 1) were prepared as described in the methods section[11] and characterized by TEM (Supplementary Figure 4). These nanoparticles are inherently covered by thiol linkers from the MPTMS precursor allowing QDs binding at the next step. During the washing step, small amount of diluted base was utilized to avoid cascade and aggregation and to ensure a uniform binding on each SiO$_2$ sphere surface in the next step.

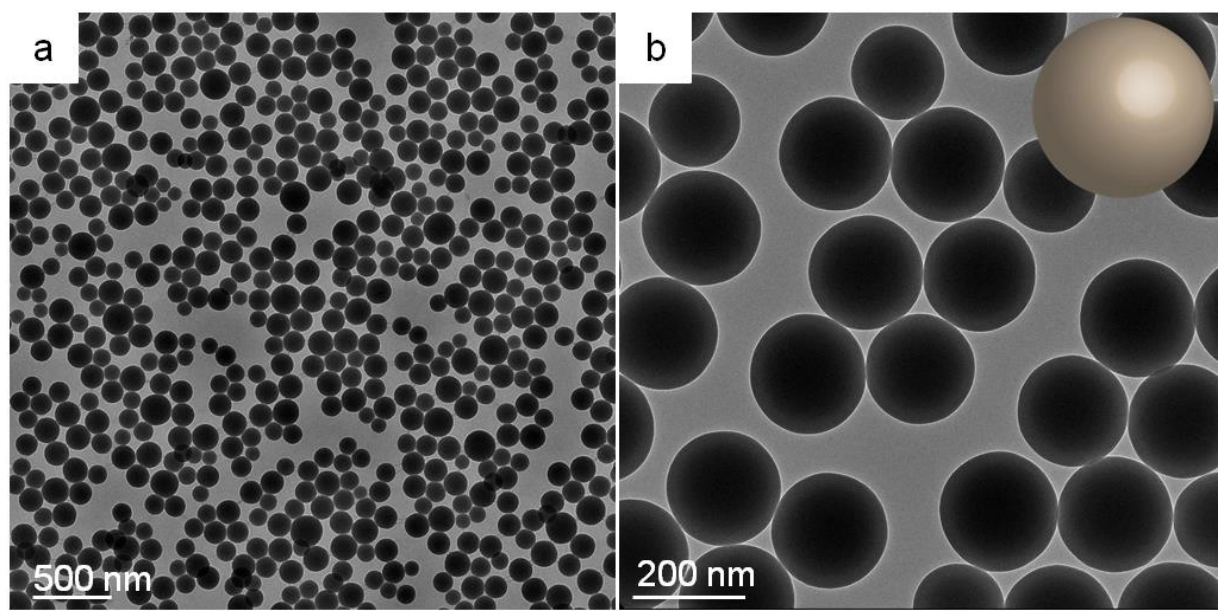

**Supplementary Figure 4.** TEM images of SiO$_2$ nanoparticles prepared by MPTMS precursor acquired at different magnifications.

**Step II - Characterization of the Silica-QDs conjugates**

The SiO$_2$@QDs particles were prepared in step 2 by adding the CdSe/CdS CQDs to the SiO$_2$ nanoparticles solution. The resulting particles were characterized by TEM and SEM as shown in Supplementary Figure 5. In order to avoid the CQDs overlap and aggregation on the SiO$_2$ surface, the ratio of CQDs added to the SiO$_2$ nanoparticles was controlled. In the sample below, a 1:500 SiO$_2$:QD ratio yielded well-separated and clearly resolved surface distribution of CQDs. The CdSe/CdS@SiO$_2$ nanoparticles solution was cleaned twice from free and weakly bound CQDs by centrifugation, discarding the supernatant and re-dispersion in toluene.

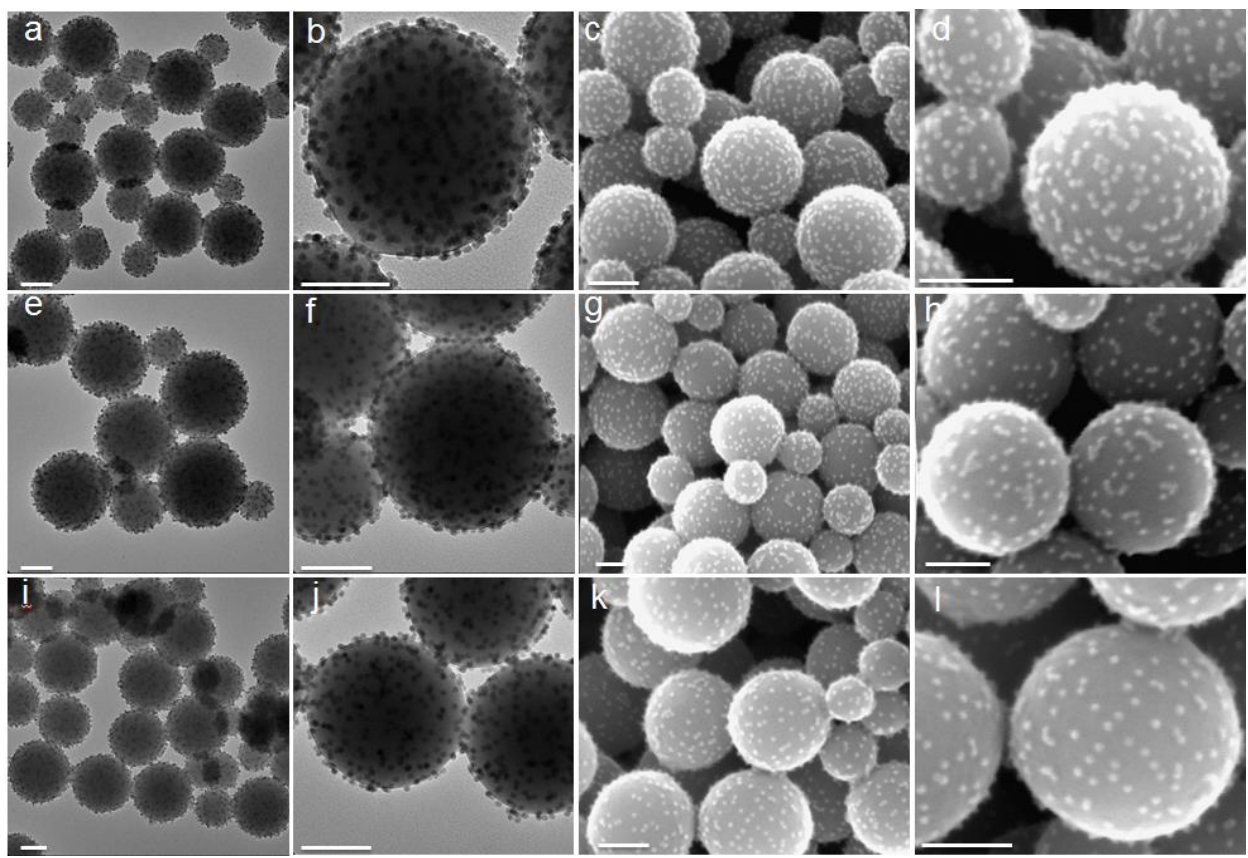

**Supplementary Figure 5. CdSe/CdS@SiO$_2$ characterized with electron microscopy.** TEM and SEM images acquired at different magnification of the CdSe/CdS@SiO$_2$ NPs produced with different loading ratios of (a-d) 1:2000, (e-h) 1:1000, and (i-l) 1:500. Scale bars are 100 nm.

**Step III – formation of a silica masking layer**

The secondary masking silica layer provides two functions: firstly, coverage of the inherent surface thiol groups of MPTMS in order to avoid the adsorption of additional CdSe-CdS CQDs in the dimerization of step 5.[11] This enhances the efficiency of the dimer formation versus monomers binding. Secondly, this immobilizes the CdSe-CdS CQDs such that they cannot rotate, exposing a hemisphere which emerges in the solvent and can be modified selectively by the chemical grafting of a functional structure/group.

In order to control the thickness of the secondary $SiO_2$ masking layer, an optimized mixture of PVP and TEOS was necessary. If the PVP amount is too low, it usually results in inefficient masking. Conversely, if it is too high it could lead to a full QDs masking by the $SiO_2$ layer, which prevents the dimer formation in the next step. Additionally, the optimization of the TEOS amount was performed as described below. According to the calculation, 50 µL of TEOS leads to a hemisphere mask. Three different amounts of TEOS were tested in our system, 200 µL (Supplementary Figure 6a-d), 100 µL (Supplementary Figure 6e-h), and 50 µL (Supplementary Figure 6i-l), respectively. For the highest TEOS loading, the CQDs were fully embedded inside the $SiO_2$ layer, as is clearly demonstrated in Supplementary Figure 6a-d, which is obviously useless for the next binding procedure. Instead, 50 µL of TEOS were utilized for the secondary essential minimal-limitation masking as well as providing the immobilization to ensure the dimer formation yield (Supplementary Figure 6i-l). After the growth of the secondary $SiO_2$ layer, the surface roughness of the CdSe-CdS@$SiO_2$ nanoparticles increases significantly, while the emergent CQDs can still be discerned. The TEM and SEM characterization of the resulting nanoparticles are presented in Supplementary Figure 6.

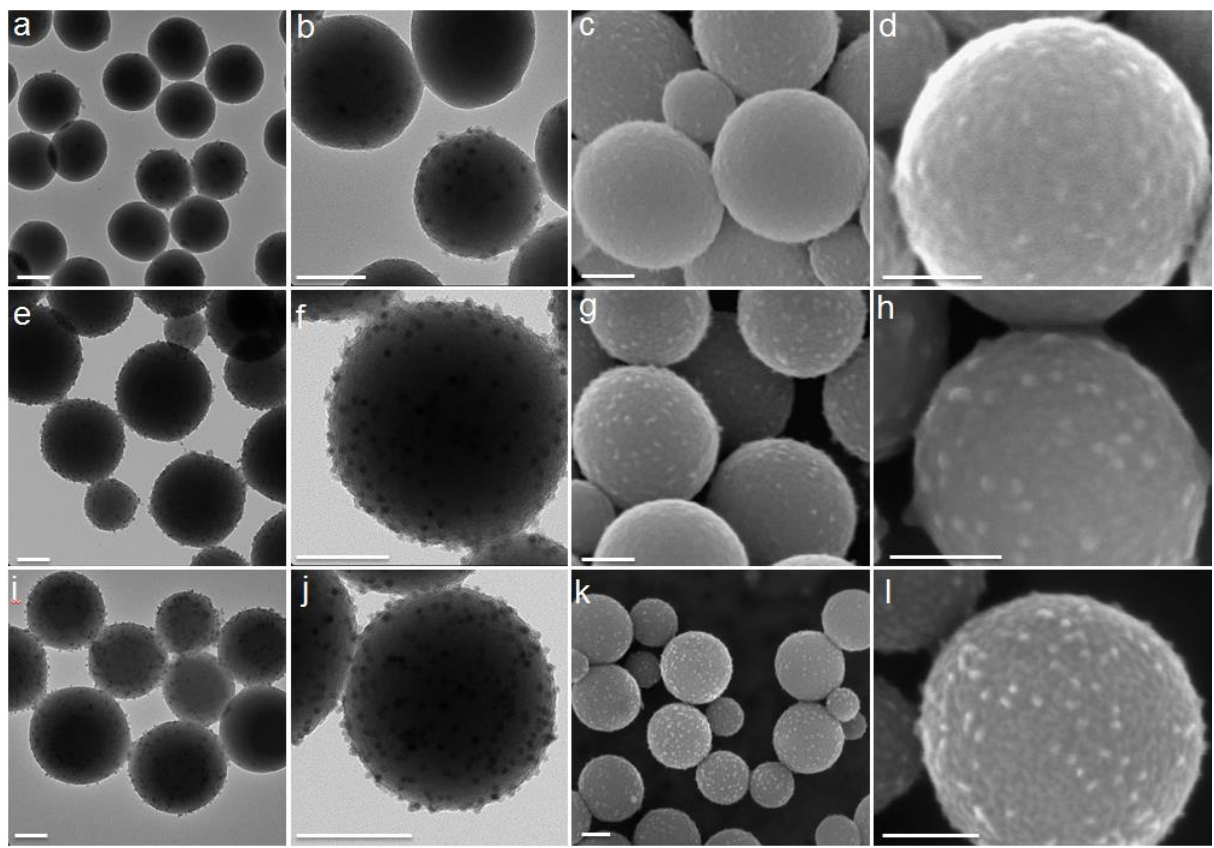

**Supplementary Figure 6.** TEM and SEM images acquired at different magnification of the SiO$_2$@CdSe/CdS@SiO$_2$ NPs produced with different amount of TEOS for masking (a-d) 200, (e-h) 100, and (i-l) 50 µL. Scale bars are 100 nm.

**Step IV – linker binding**

For the dimer formation step, the chosen linkers bind to the exposed region of the anchored CQDs (step 4). A Tetra-thiol molecule was used as a bi-dentate linker molecule (Supplementary Figure 7). The thiol groups strongly bind to the CQD surface and can displace the existing surface ligands of the exposed CQD hemisphere. In order to enhance the conjugation of the linkers, the surface modification procedure was performed under Argon flow, at 60 °C overnight. The excess linker molecules were removed by precipitation and centrifugation. Here, the cleaning step after the linker addition was significant for achieving high dimer formation yield.

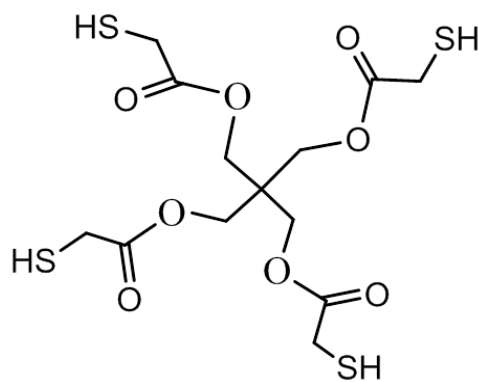

**Supplementary Figure 7.** Chemical structure of the linker (pentaerythritol tetrakis(3-mercapto-propionate)).

**Step V – Dimer formation**

For the preparation of homodimers a CQDs, a ratio of 1:1.2 to the original amount used in step 2 was employed. The resulted structures are presented in Supplementary Figure 8.

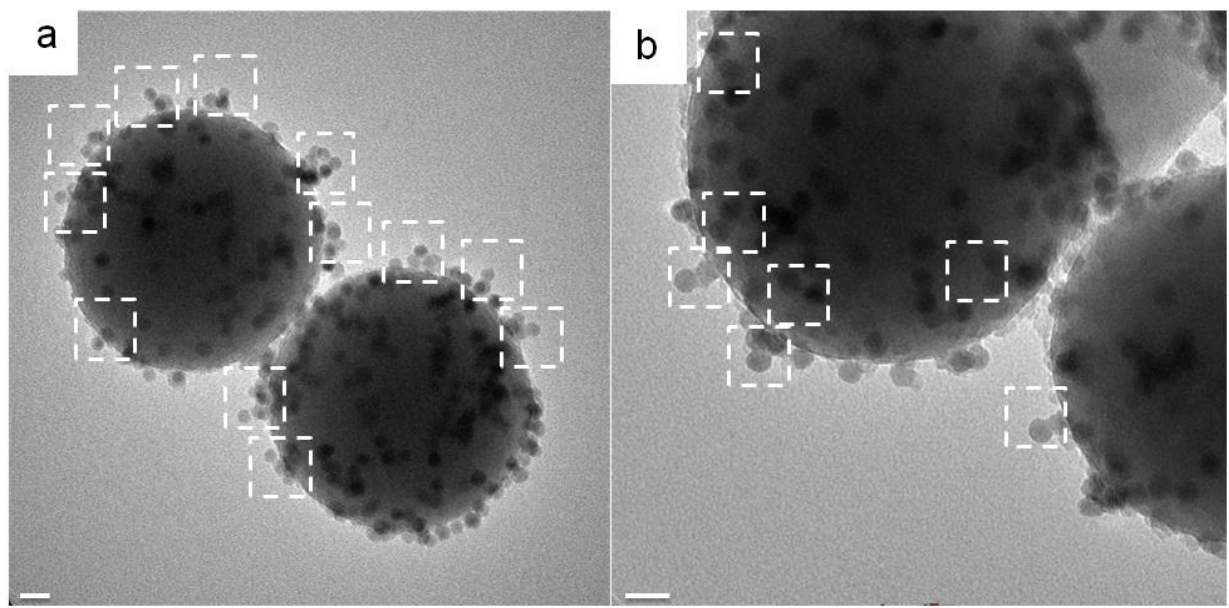

**Supplementary Figure 8.** TEM images of SiO$_2$@dimer CdSe/CdS particles acquired at different magnifications. Scale bars are 20 nm. Representative dimer structures are marked by white frames.

**Step VI – Dimers release**

Precisely controlled dimer CdSe-CdS CQD molecules were successfully achieved as follows. The release and separation of the CdSe-CdS CQDs dimers from the SiO$_2$ spheres was performed by selective etching process of the SiO$_2$ using an HF/NMF (10%) etching solution. The free dimers are shown in Fig. 1c. The freed dimers were separated by centrifugation decanting the supernatant and repeated ethanol precipitation/centrifugation cycles for three times.

**Step VII – Fusion to form the coupled NC molecule**

The fusion procedure plays a significant role in reduction of the potential barrier in the coupled CQDs molecules. The choice of correct temperature, precursor amount and ligands lead to the fusion of two CQDs without ripening and collapse.

**Size selective separation process for 1.9/4.0 nm CQDs dimers**

The fused dimer molecules as prepared by the procedure explained in the previous section contain some unreacted monomer CQDs which have been purified to achieve high yield of CQD molecule for further studies. Supplementary Figure 9 shows the released fused dimers, which were separated by centrifugation decanting the supernatant and repeated redispersion (ethanol)/precipitation (centrifugation) cycles for three times. Monomers were found mainly in the suspension, as shown in Supplementary Figure 9b. As seen in Supplementary Figure 9c, the monomers, along with possible higher order aggregates of CQDs can be well separated after this step, resulting in a purified solution of dimer structures.

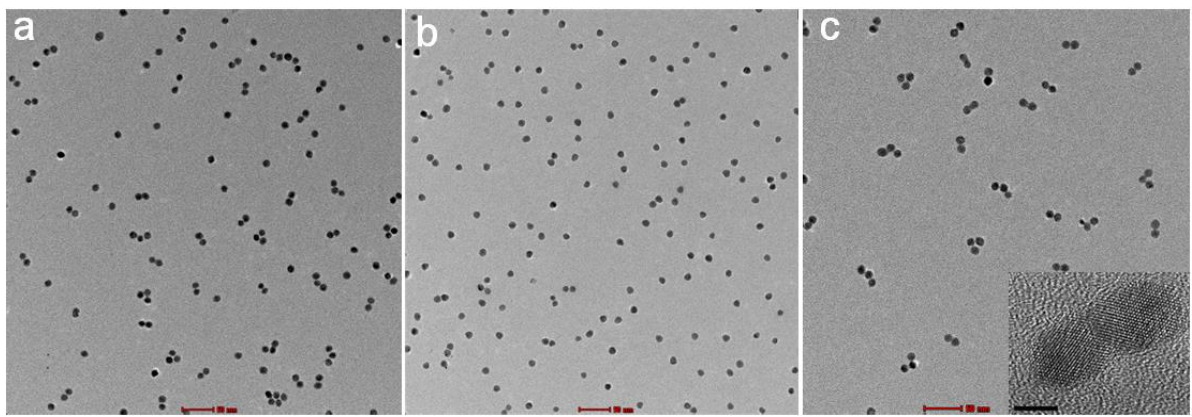

**Supplementary Figure 9.** TEM images of fused 1.9/4.0 nm CdSe/CdS molecules before size selective separation process (a). The suspension (b) and the precipitate (c) of fused CdSe/CdS molecules after the size selective separation process. Inset, HRTEM image of fused CdSe/CdS molecules.

**Size selective separation process for 1.4/2.1 nm CQDs dimers**

The size selection separation for other nanocrystal sizes is similar but may require slight modifications. For example, for 1.4/2.1 nm CdSe/CdS CQD molecules low centrifugation rate was used for efficient separation. The resulting dimers are presented in Supplementary Figure 10.

After the separation step, the sample was stored under argon atmosphere and kept in the glovebox for further characterization and single nanoparticle measurement.

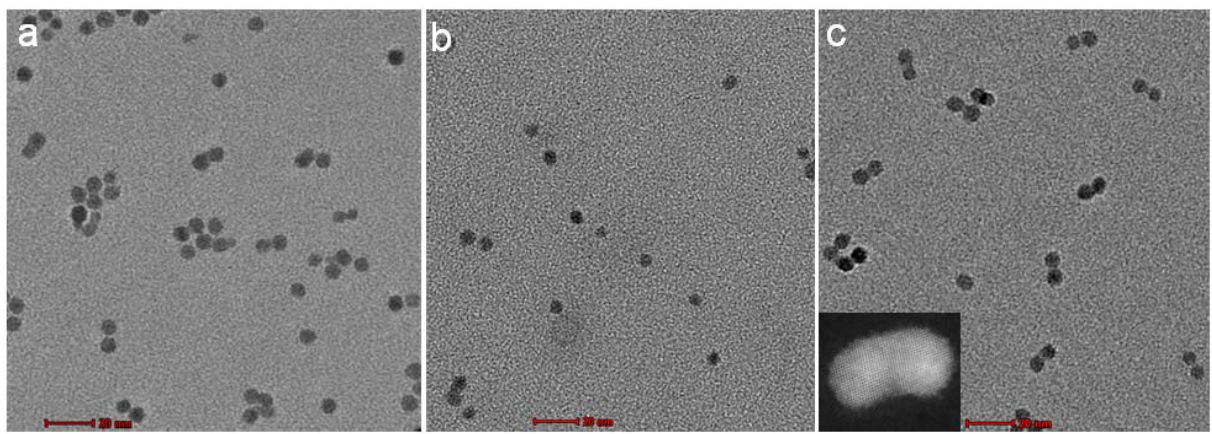

**Supplementary Figure 10.** TEM images of fused 1.4/2.1 nm CdSe/CdS molecules (a) before size selective separation process. The suspension (b) and the precipitate (c) of fused CdSe/CdS molecules after the size selective separation progress. Inset, HRTEM image of fused CdSe/CdS CQD molecule.

**Absorption spectrum measurement for 1.4/2.1 nm CQDs**

The features in the absorption spectrum change significantly after the fusion of two CQDs. To further demonstrate the changes, we have normalized the data at the bulk absorption regime (300nm). It can be observed that the band-edge absorption feature of the unfused dimer is retained and similar to the monomers. Upon fusion broadening is seen, along with a red shift and lower absorption. The excited state features are also broadened significantly upon fusion. Upon normalization at the band-edge, the significant relative change in absorption at higher energies upon fusion is emphasized (Supplementary Figure 11a). All these aspects indicate coupling effects upon fusion and formation of the CQD molecules.

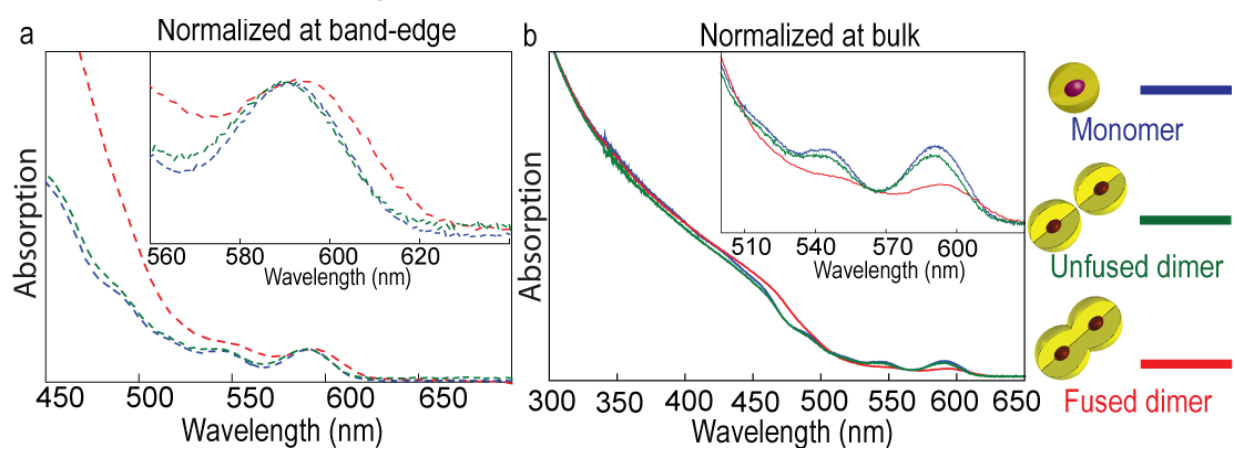

**Supplementary Figure 11.** The normalized absorption spectra of monomers (blue), unfused (green), and fused 1.4/2.1 nm CdSe/CdS CQD molecules when normalized with respect to (a) band-edge peak and (b) bulk transitions (300 nm).

**Control experiment: monomer particles that were treated with the fusion procedure**

The CQDs are exposed to different conditions during the synthetic procedure such as binding, etching, fusion etc. An important control is to identify whether the inherent properties of the CQDs change during these processes, especially during etching and fusion. Hence, we studied the absorption and photoluminescence spectra (Supplementary Figure 12) of monomers that although did not form a dimer, were still treated with the complete synthetic procedures of the fusion protocol. These monomers were separated by size selective precipitation (Supplementary Figure 10 b) and measured accordingly. The identical spectral features for the monomers, that underwent the fusion procedure, and the original ones unequivocally confirms no change in the excitonic behavior of the CQDs. This rules out the possibilities of any shift due to alloying between core/shell regions.

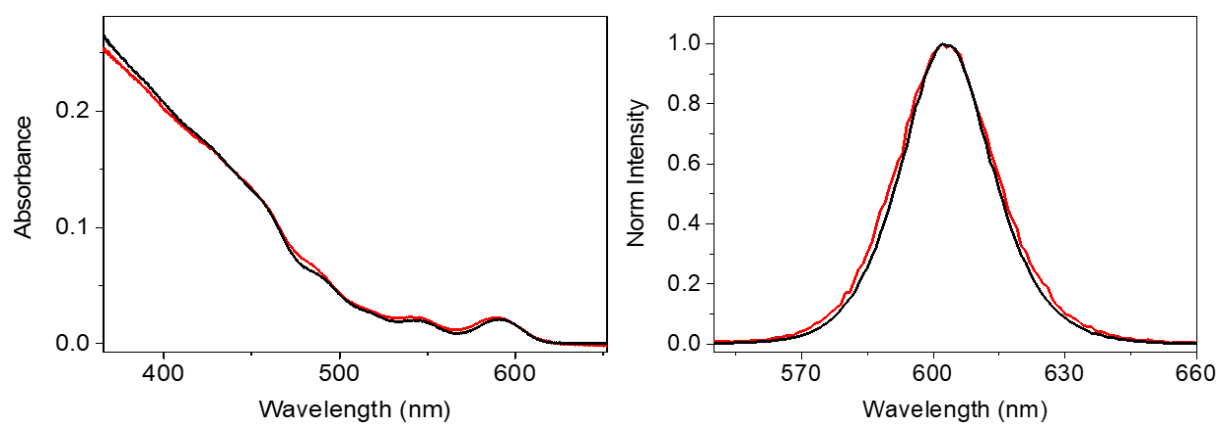

**Supplementary Figure 12.** The ensemble absorption and PL spectra of monomer before (black), and after the fusion procedure (red) for CdSe-CdS CQDs with core/shell size of 1.2/2.1 nm.

**Spectral characterization of the various fused CQDs dimers**

The red shift of the fluorescence was enhanced with a size decrease in the CdSe-CdS CQDs composing the fused dimers. That is, the smaller core-shell CQDs molecules present strong coupling properties. Additionally, upon fusion, the full width at half maximum (FWHM) measured from the fluorescence spectrum was obviously increased compared with that of the monomer sample.

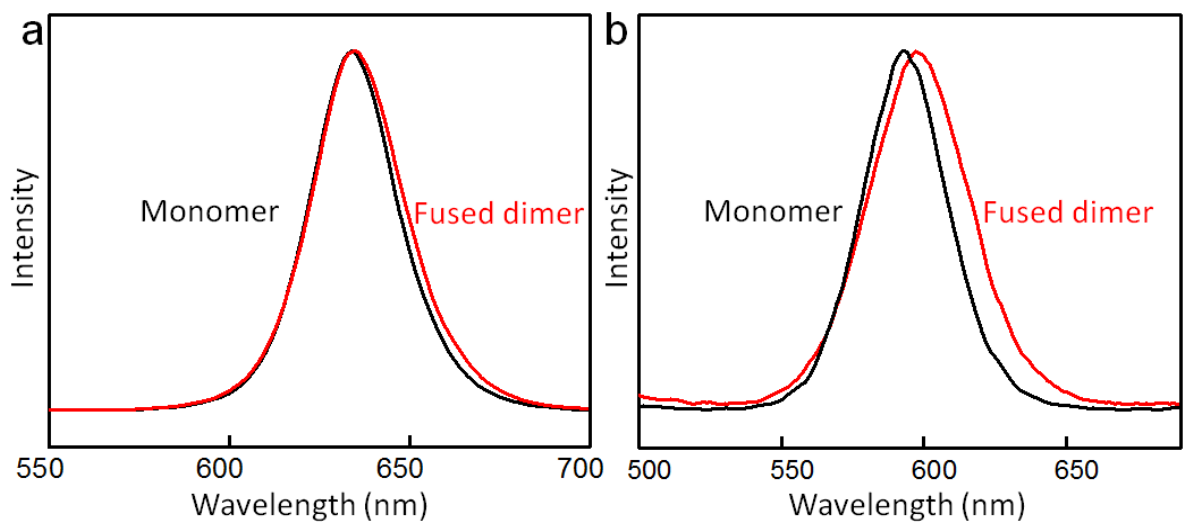

**Supplementary Figure 13.** The PL intensity of monomer (black), and fused (red) CdSe-CdS CQDs molecules with core/shell size of 1.9/4.0 nm (a), and 1.2/2.1 nm (b), respectively.

**Structural characterization of the fused 1.9/4.0 nm CQDs dimers**

Supplementary Figure 12 shows HAADF-STEM images and atomic structure models of the fused 1.9/4.0 nm CdSe/CdS CQD molecules with hetero-plane orientation relationship at attachment. The analysis allowed to determine the (0002) faces (and thus C-axis orientation) for each CQD in a pair as indicated on the images and structure models in Supplementary Figure 14. It is clearly seen that the C-axes of the monomers in a heteronymous-plane fused dimer are randomly oriented unlike the distinct parallel alignment observed at homonymous-plane attachment (see main text).

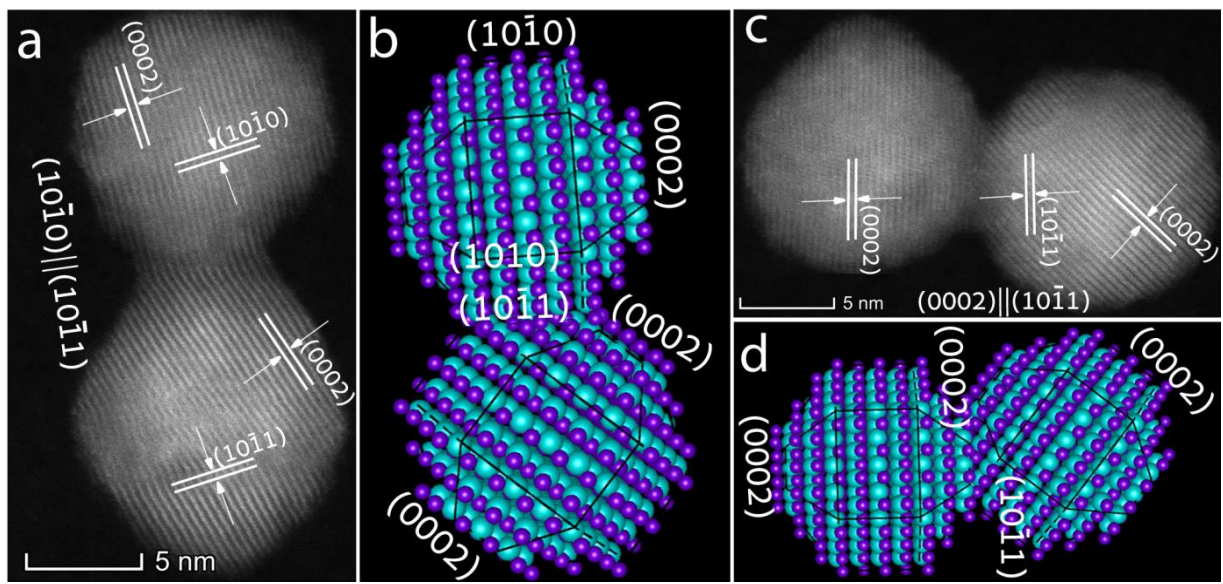

**Supplementary Figure 14.** HAADF-STEM images and atomic structure models of the fused 1.9/4.0 nm CdSe/CdS molecules with hetero-plane attachment of $(10\bar{1}0)\|(10\bar{1}1)$ (a-b), $(0002)\|(10\bar{1}1)$ (c-d). For the atomic model, the Cadmium atoms are marked in brown and Sulfur atoms in blue.

**Structural characterization of the fused 1.4/2.1 nm CQDs dimers**

In our system, the attachment and fusion of the monomer CdSe/CdS CQDs was based on the linker's binding. Therefore, the attachment orientation relationships for the smaller CdSe/CdS CQDs molecules were similar to those observed for the larger CQDs. Yet, a small difference was still noticed. Namely, there are three main options for homonymous-plane attachment: (0002)∥(0002), (10$\bar{1}$0)∥(10$\bar{1}$0), and (10$\bar{1}$1)∥(10$\bar{1}$1). Supplementary Figure 13a shows the (10$\bar{1}$0)∥(10$\bar{1}$0) homonymous-plane attachment with (10$\bar{1}$0) faces of A1 parallel to those of A2 and a continuous lattice observed through the entire dimer.

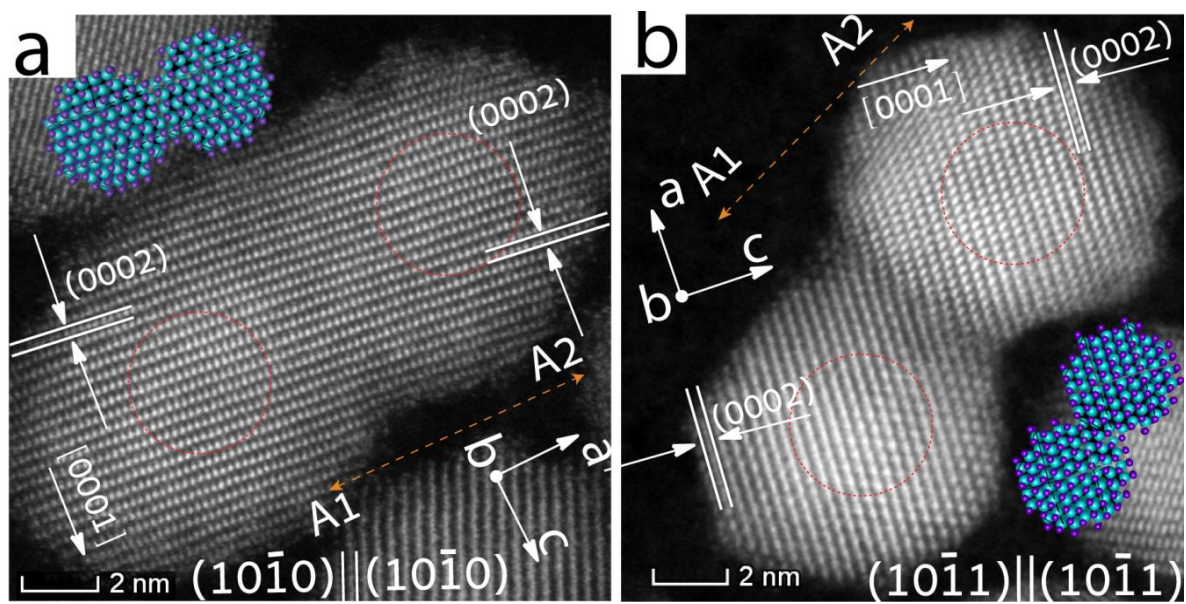

**Supplementary Figure 15.** HAADF-STEM image and atomic structure model of the fused 1.4/2.1 nm CdSe/CdS molecules with homo-plane attachment on (10$\bar{1}$0)∥(10$\bar{1}$0) (a), and (10$\bar{1}$1)∥(10$\bar{1}$1) (b). For the atomic model, the Cadmium atoms are marked in purple and Sulfur atoms in light blue.

**Heteronymous-plane attachment of the fused 1.4/2.1 nm CQDs dimers**

The hetero-plane attachment mainly based on $(0002)\|(10\bar{1}0)$, $(0002)\|(10\bar{1}1)$, and $(10\bar{1}1)\|(10\bar{1}0)$ was also observed in the fused 1.4/2.1 nm CdSe/CdS CQD molecules. In this case, due to the thin CdS shell (only 6 atomic layers thick), the core CdSe region (Cd: atom with high brightness, Se: atom with slight brightness) was directly identified at HAADF STEM image. Supplementary Figure 14 shows examples of heteronymous-plane attachments at $(0002)\|(10\bar{1}0)$ (a) and $(0002)\|(10\bar{1}1)$ (b).

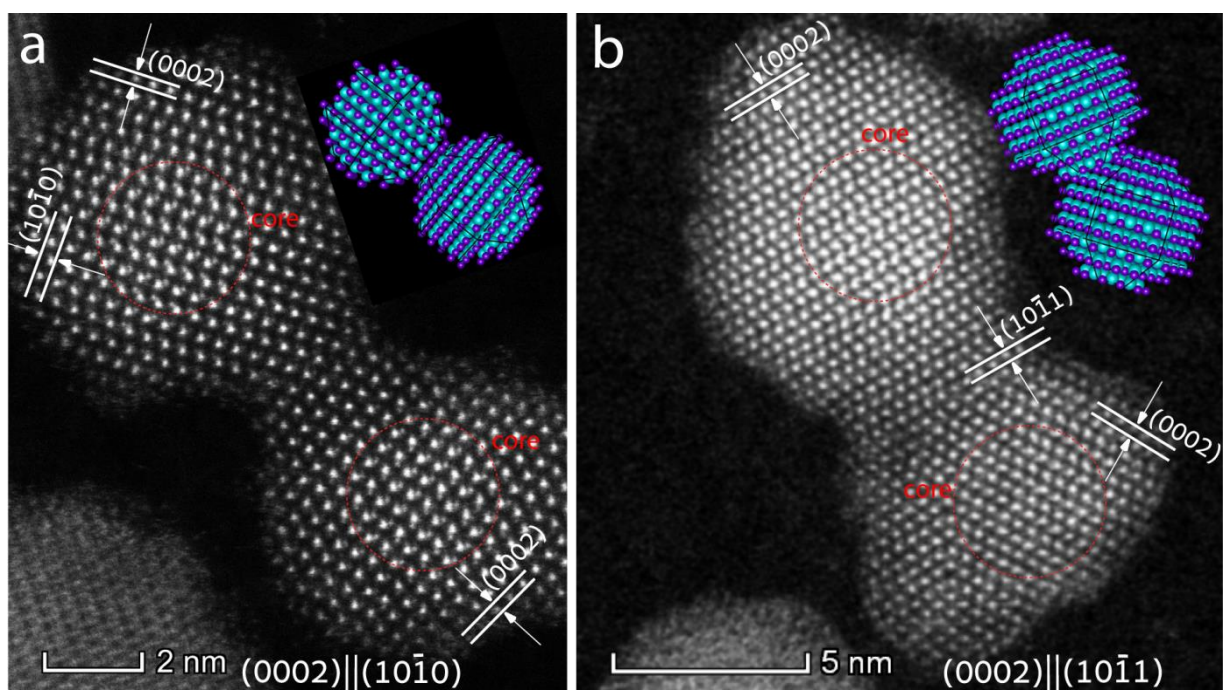

**Supplementary Figure 16.** HAADF-STEM images and atomic structure model of the fused 1.4/2.1 nm CdSe/CdS molecules with heteronymous-plane attachment on $(0002)\|(10\bar{1}0)$ (a), $(0002)\|(10\bar{1}1)$ (b). For the atomic model, the Cadmium atoms are marked in purple and Sulfur atoms in light blue.

**Ensemble fluorescence lifetime measurements for 1.9/4.0 nm CQDs and the corresponding dimers**

The molecule formed from larger CQDs (1.9/4.0 nm) also exhibit a quenching in the ensemble lifetime upon dimer formation but at a much lower extent compared with molecules of the small CQDs (1.4/2.1 nm). This is due to reduced tunneling and hybridization both because of the longer barrier and also due to the larger cores with less wavefunction spilling into the shell.

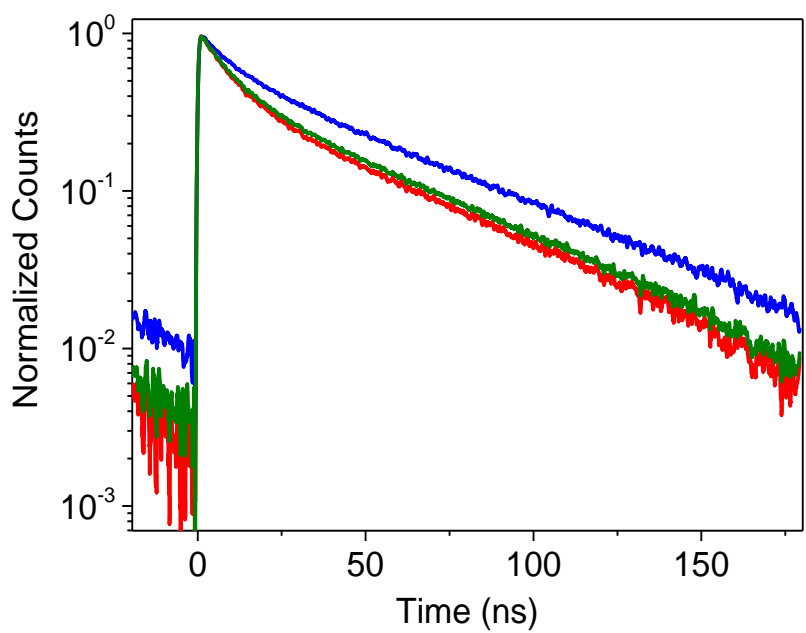

**Supplementary Figure 17.** Ensemble fluorescence lifetime decay curves of the monomer (blue), unfused dimer (green) and fused dimer (red) for 1.9/4.0 nm CQDs.

**Single particle fluorescence lifetime for small 1.4/2.1 nm CQDs and the corresponding dimers**

The fluorescence decay of a single 1.4/2.1 nm CQD follows mostly a mono-exponential decay of ~30ns or a bi-exponential decay with a small contribution from a 5 ns component. The average lifetime for the single dimer particles is clearly quenched and follows mostly a tri-exponential decay. The distribution of the average lifetime for dimer particles is shifted to short values, which is more pronounced for the fused dimers than for the unfused ones.

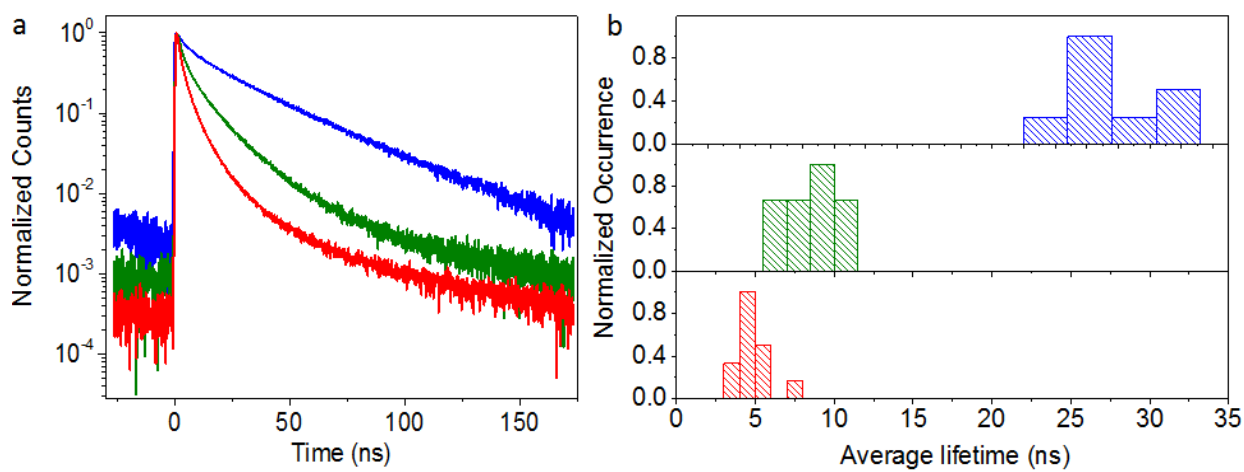

**Supplementary Figure 18.** Single particle fluorescence lifetime decay data for monomer (blue), unfused dimer (green) and fused dimer (red) composed of 1.4/2.1 nm CQD. (a) Representative single particle fluorescence lifetime decays. (b) Histograms summarizing the distribution of the average lifetimes for all three types of particles. Further shortening of the lifetime is observed upon fusion.

**Characterization of the monomer particles that were treated with the fusion procedure:**

The fraction of monomer particles found is in accordance with the size selective precipitation results and TEM analysis therein (~15%, Supplementary Figure 10). The optical properties of single monomer CQDs (Supplementary Figure 19-20) that although did not form a dimer, still were treated with the complete synthetic procedures of the fusion protocol, were characterized. We found that the monomer CQDs undergoing the fusion protocol retain their optical properties exactly as the unprocessed monomers, such as single exponential lifetime of the on state, on-off blinking of fluorescence and strong photon antibunching.

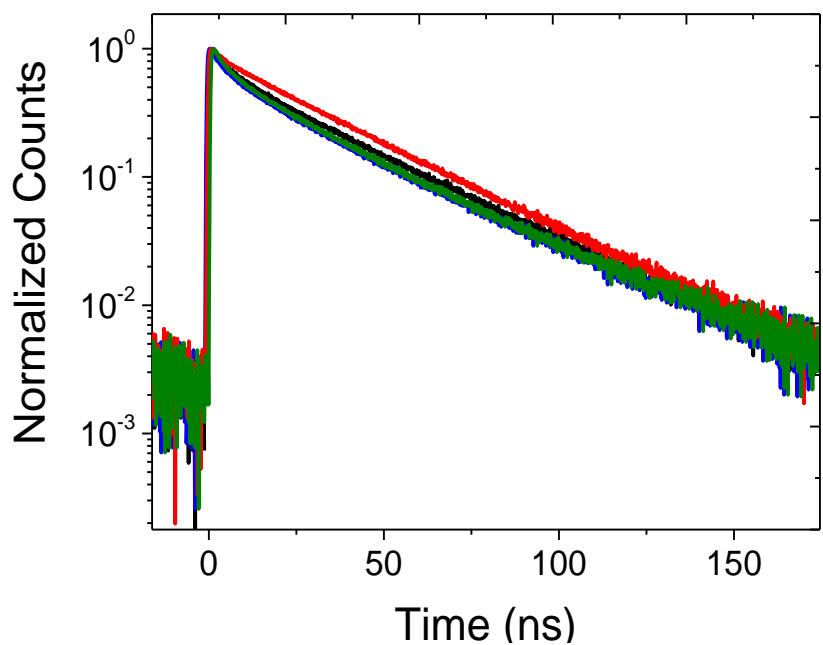

**Supplementary Figure 19**. Single particle fluorescence lifetime of a monomer CQD (1.4/2.1 nm) found along with the fused dimers. Two examples of the fusion protocol treated monomers are shown in blue and red curves. The black and green curves represent the lifetime of monomer CQDs that was not treated with the fusion procedure. The deviations of all representative average lifetimes are well under the distribution shown in Supplementary Figure 18.

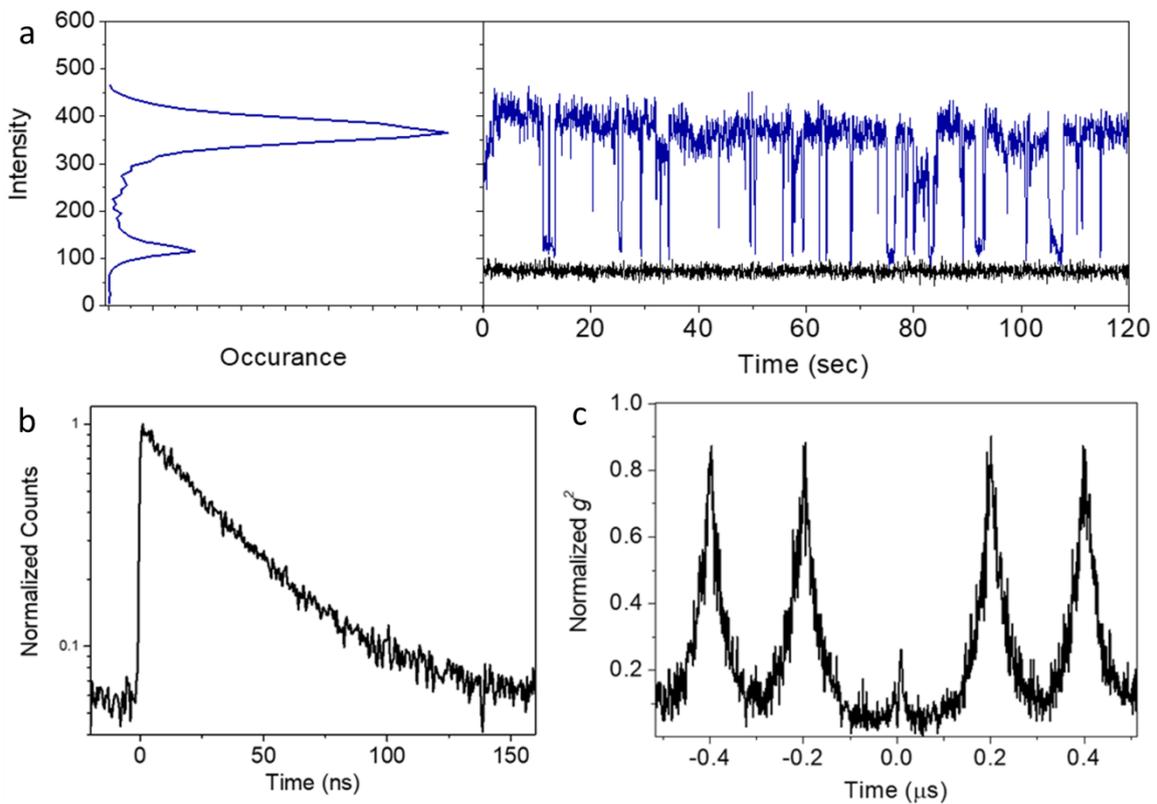

**Supplementary Figure 20.** Time-tagged time-resolved analysis of the fusion protocol treated monomer CQD (1.4/2.1 nm). (a) A bimodal on-off distribution of the intensity was found (bin-50 ms). The black curve represents the background noise. (b) Fluorescence lifetime of the on-state follows a single exponential decay of 31 ns. (c) Strong photon antibunching with $g^2$ value of 0.09. All these observations for the fusion protocol treated monomer are highly correlated with the untreated monomer particle as explained in the previous section.

**Single particle fluorescence lifetime for 1.9/4.0 nm CQDs and the corresponding dimers**

The shortening of the lifetime upon dimerization was also observed for the 1.9/4.0 nm CQDs. When compared with the lifetime distribution of the smaller particles (Supplementary Figure 18), the degree of shortening was found to be lower in this case. A similar type of distribution in the averaged lifetime for unfused and fused dimer indicates absence of significant additional coupling upon fusion, consistent with decreased tunneling-coupling in these molecules constructed from the larger CQDs.

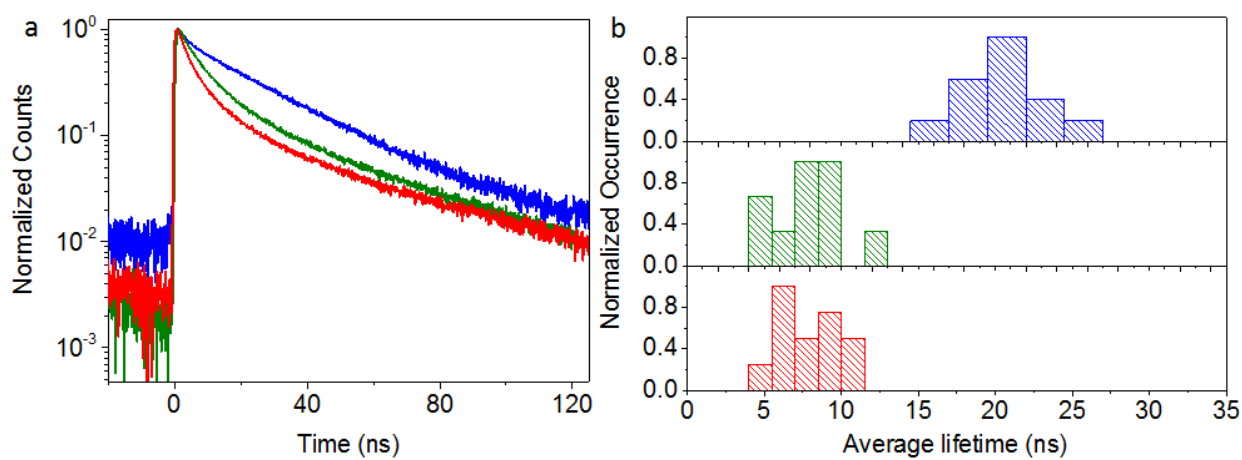

**Supplementary Figure 21.** Single particle fluorescence lifetime decay data for monomer (blue), unfused dimer (green) and fused dimer (red) composed of 1.9/4.0 nm CQD. (a) Representative single particle fluorescence lifetime decays (b) Histograms summarizing the distribution of average lifetimes for all three types of particles. For the shake of comparison, x-axis was kept the same as in Supplementary Figure 18.

**Comparison of g² value for monomer and fused dimer**

The antibunching at zero time delay accounts for the ratio of biexciton to exciton quantum yield (QY) $g^2 = \frac{Area_{(0ns)}}{Area_{(200\,ns)}} = \frac{QY_{(BX)}}{QY_{(X)}}$ when the value of <N> is kept well below 1. When the monomer (1.4/2.1 nm) and fused dimer were excited with similar <N> (~0.06) value, a much higher value of g² was obtained in the case of the fused dimer (0.68) than for the monomer (0.16) CQD. The possible pathways for the enhanced biexciton QY are explained in the main text.

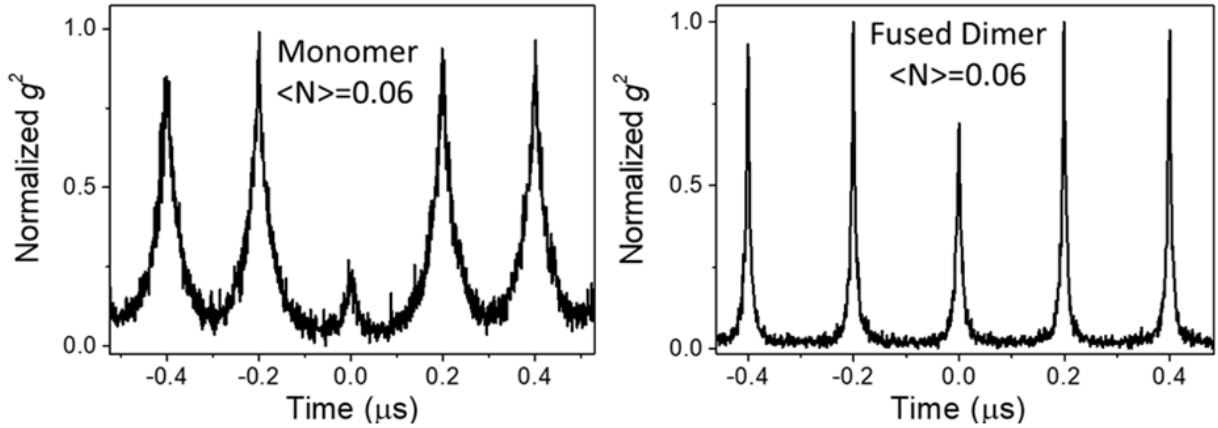

**Supplementary Figure 22.** Comparison of second order photon correlation at similar <N> value for excitation. The g² value were found to be 0.16 and 0.68 for the monomer and the fused dimer, respectively.

**Attributes of fluorescence from single unfused dimer**

The fluorescence from single unfused dimers also exhibited flickering nature instead of distinct on-off characteristics and followed a multi-exponential decay. The lifetime is not uniform throughout the intensity range, but rather presents distributions when analyzed at different intensity levels (Supplementary Figure 23(ii)). The single unfused dimer also gives rise to lower antibunching contrast.

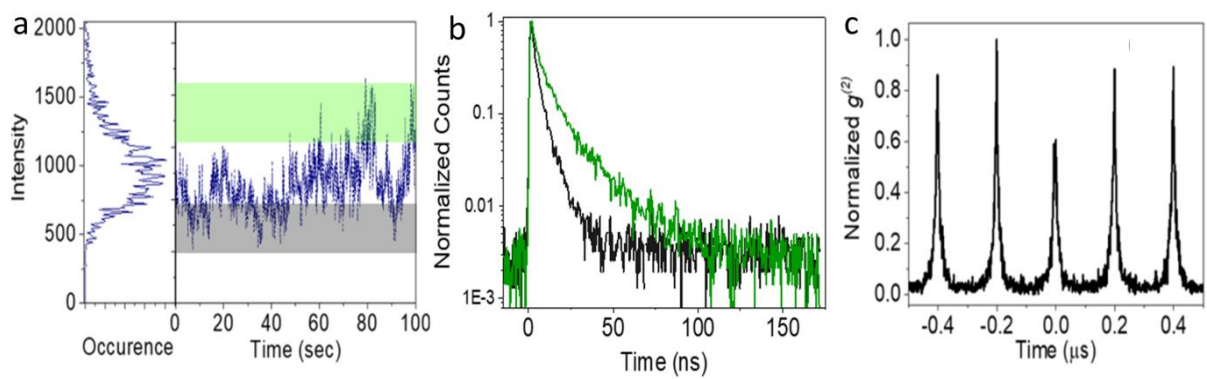

**Supplementary Figure 23**. Time-tagged time-resolved data for single unfused dimer comprised of 1.4/2.1 nm CQD(a) Photoluminescence time trace, (b) fluorescence intensity dependent lifetime (the lifetime curves bear the same color code as the shaded area from (a)), and (c) second order photon correlation ($g^2$).

**Excitation intensity dependence of fluorescence flickering in dimer**

The flickering nature of the fluorescence from a single fused dimer persists even in very low excitation conditions. The brown colored time trace in Supplementary Figure 24 was obtained in the lowest possible excitation (<N>~0.03), where the flickering of fluorescence occurs without reaching an off state (black curve). With increase in the excitation intensity an increment of the extent of flickering was observed along with shortening of the lifetime as described in Fig. 5f in the main manuscript.

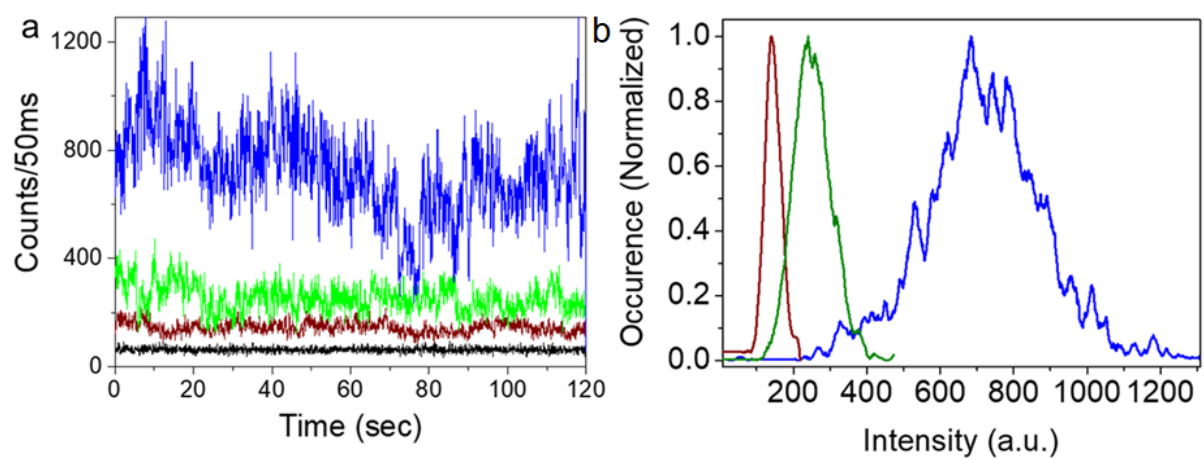

**Supplementary Figure 24.** Fluorescence time traces information with varying excitation intensity for a fused dimer comprised of 1.4/2.1 nm CQDs. The blue, green and brown curve represents the traces obtained from fused dimer with <N> value of 0.18, 0.09, 0.03, respectively. (a) Excitation fluence dependent PL time traces. The black curve represents the background noise in a similar configuration. (b) Intensity distribution of the traces in (a). Narrow distribution is observed with decreasing laser power.

**Supplementary Table 2.** The calculated and experimental monomer-to-dimer red shift for CdSe/CdS CQD molecules with different core/shell diameters.

| Core/Shell (nm) | Monomer emission peak (nm) | Fused emission peak (nm) | Red-shift (meV)-Exp | Red-shift (meV)-Calc |
|---|---|---|---|---|
| **1.9/4.0** | 637 | 637 | 0 | 0.65 |
| **1.4/2.1** | 603 | 606.5 | 11.8 | 11.3 |
| **1.2/2.1** | 593 | 597 | 13.9 | 12.8 |

**Supplementary Table 3**. The ensemble fluorescence lifetime values for different types of CQDs and corresponding homodimers (in toluene).

| Core/Shell (nm) | Monomer | | | Unfused Dimer | | | Fused Dimer | | |
|---|---|---|---|---|---|---|---|---|---|
| | $\tau_1(a_1)$ | $\tau_2(a_2)$ | $\tau_{avg}$ | $\tau_1(a_1)$ | $\tau_2(a_2)$ | $\tau_{avg}$ | $\tau_1(a_1)$ | $\tau_2(a_2)$ | $\tau_{avg}$ |
| 1.4/2.1 | | 25 (1.0) | 25 | 7.3 (0.9) | 30 (0.1) | 8.7 | 5.2 (0.3) | 9.7 (0.7) | 9.7 |
| 1.9/4.0 | 8.9 (0.4) | 39 (0.6) | 24.5 | 7.9 (0.62) | 38 (0.38) | 17.5 | 8.1 (0.65) | 39 (0.35) | 17.1 |

$\tau_i$ values are represented in ns scale, $a_i$ represents the weightage of a particular lifetime component

**Supplementary References**